\documentclass{aa}

\usepackage{graphicx}
\usepackage{txfonts}
\usepackage{hyperref}
\usepackage{subcaption}
\usepackage{tabularx}
\usepackage{ulem}
\defcitealias{libra2015}{Paper\,I}

\begin{document}

\title{High-precision astrometry with VVV}

   \subtitle{II. A near-infrared extension of \textit{Gaia} into the Galactic plane}

   \author{M. Griggio\inst{1,2,3}
          \and
          M. Libralato\inst{1,4}
          \and
          A. Bellini\inst{3}
          \and
          L. R. Bedin\inst{1}
          \and
          J. Anderson\inst{3}
          \and
          L. C. Smith\inst{5}
          \and
          D. Minniti\inst{6,7,8}
          }

   \institute{INAF - Osservatorio Astronomico di Padova, Vicolo
               dell'Osservatorio 5, Padova I-35122, Italy
               \and
               Dipartimento di Fisica, Universit\`a di Ferrara, Via Giuseppe Saragat 1,
               Ferrara I-44122, Italy
               \and
               Space Telescope Science Institute, 
               3700 San Martin Drive, Baltimore, MD 21218, USA
               \and
               AURA for the European Space Agency, Space Telescope Science Institute, 3700 San Martin Drive, Baltimore, MD 21218, USA
               \and
               Institute of Astronomy, University of Cambridge, Madingley Rd, Cambridge CB3 0HA, UK
               \and
               Institute of Astrophysics, Universidad Andres Bello, Fernandez Concha 700, Las Condes, Santiago, Chile
               \and
               Vatican Observatory, V00120 Vatican City State, Italy
               \and
               Departamento de Física, Universidade Federal de Santa Catarina, Trindade 88040-900, Florianópolis, Brazil\\
               \email{massimo.griggio@inaf.it}
             }
    
   \date{Received 12 February 2024; accepted 18 March 2024}

 
  \abstract
   {}
   {We use near-infrared, ground-based data from the {\it VISTA Variables in the Via Lactea} (VVV) survey to indirectly extend the astrometry provided by the {\it Gaia} catalog to objects in heavily-extincted regions towards the Galactic bulge and plane that are beyond {\it Gaia}'s reach.}
   {We make use of the state-of-the-art techniques developed for high-precision astrometry and photometry with the {\it Hubble Space Telescope} to process the VVV data. We employ empirical, spatially-variable, effective point-spread functions and local transformations to mitigate the effects of systematic errors, like residual geometric distortion and image motion, and to improve measurements in crowded fields and for faint stars. We also anchor our astrometry to the absolute reference frame of the {\it Gaia} Data Release 3.}
   {We measure between 20 and 60 times more sources than {\it Gaia} in the region surrounding the Galactic center, obtaining an single-exposure precision of about 12\,mas and a proper-motion precision 
   of better than 1\,mas\,yr$^{-1}$ for bright, unsaturated sources.
   Our astrometry provides an extension of {\it Gaia} into
   the Galactic center. We publicly release the astro-photometric catalogs of the two VVV fields considered in this work, which contain a total of $\sim$\,3.5\,million sources. Our catalogs cover $\sim$\,3\,sq.\,degrees, about 0.5\% of the entire VVV survey area.}
   {}


   \titlerunning{An extension of \textit{Gaia} into the Galactic plane with VVV}
   \authorrunning{M. Griggio et al.}

   \maketitle

\section{Introduction}

\begin{figure*}
    \centering
    \includegraphics[width=\textwidth]{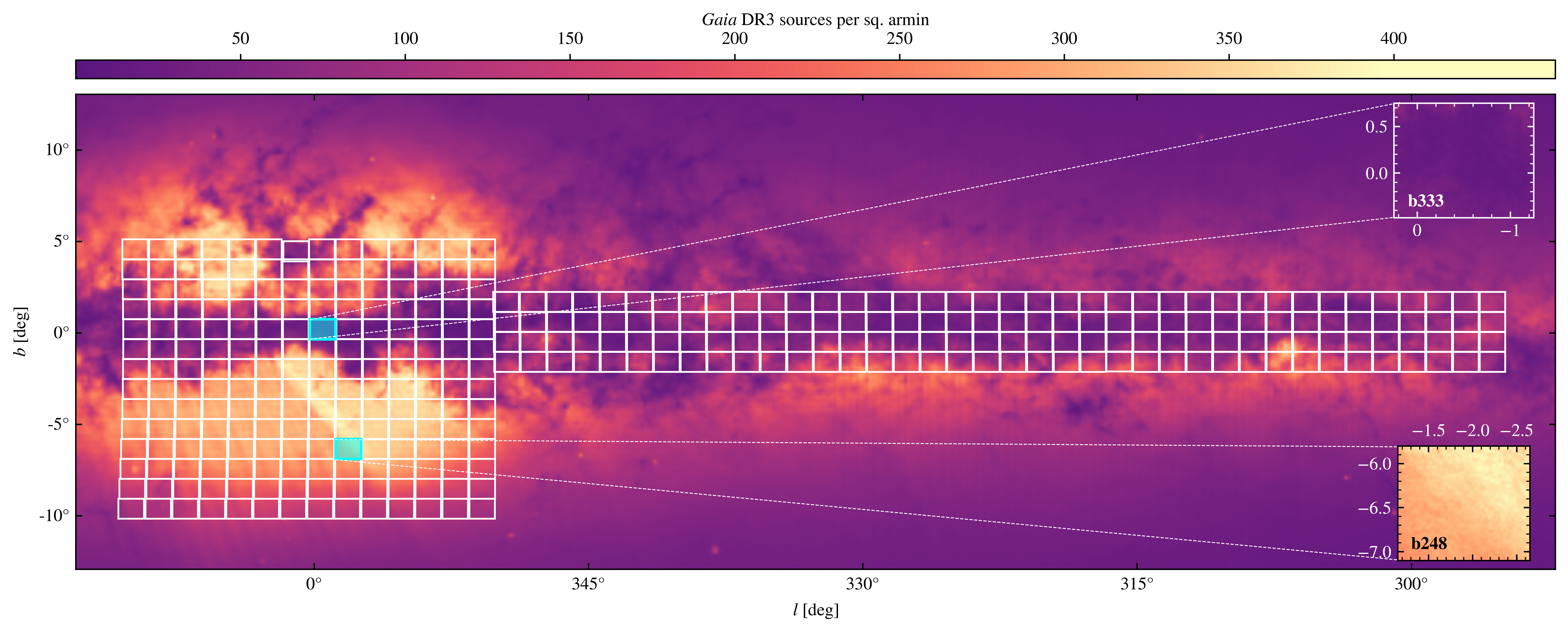}
    \caption{{\it Gaia} DR3 source density around the region
    covered by the VVV survey. White boxes represent VVV tiles. We highlighted in light blue the two tiles considered in this work,
    namely \texttt{b333} in the Galactic center, and \texttt{b248} on
    the southern Bulge. Density map data taken from the {\it Gaia} archive.}
    \label{fig:source_d}
\end{figure*}

The {\it Gaia} mission \citep{gaiamission} has revitalized many scientific research fields by providing the community with exquisite astrometry and photometry for more than a billion sources. All-sky coverage and astrometric accuracy and precision also make the {\it Gaia} catalogs among the best resources for technical projects like catalog/image registration
to the International Celestial Reference System.
The {\it Gaia} mission in fact provides an absolute astrometric reference frame on the sky, which can be used 
to derive geometric-distortion corrections for many cameras, as shown by, e.g., \cite{2022MNRAS.515.1841G}.

However, {\it Gaia} suffers from a few shortcomings, one of which being highly incomplete even for bright sources in
high-density fields, such as in the proximity of globular clusters or the Galactic bulge \citep{2021A&A...649A...5F}.
In addition, the {\it Gaia} catalog contains only sources
brighter than $G$\,$\sim$\,21,
which greatly limits its effectiveness in
dust-enshrouded regions, such as those towards the Galactic plane.
For these reasons, {\it Gaia} has limited applications for scientific or technical programs requiring high-precision
astrometry for reddened Bulge and Disk objects.

The Galactic bulge is the ideal laboratory to study stellar interactions in high-density environments on galactic scales,
providing insights into the early history of the Milky Way 
\citep[e.g.,][]{2018ARA&A..56..223B,2019BAAA...61..137Z,2020IAUGA..30..282F}, and astrometry represents a valuable tool to
investigate the stellar populations in the Bulge. In particular, proper motions allow to separate Bulge and Disk stars and to
identify gravitationally-bound systems such as comoving groups, star clusters and stellar streams
\citep[e.g.,][]{2022A&A...662A..95G,2022ApJ...940...76K}.
Moreover, the Galactic bulge is where most of the microlensing events have been discovered 
-- given the high density of sources in the Bulge -- and precise astrometry plays a fundamental role in the determination
of the geometry of the event and of the masses of the involved bodies \citep{2019ApJS..244...29M}.
Proper motion measurements have also been essential for studies of new and old globular clusters in the Milky Way
\citep{2017ApJ...849L..24M,2021A&A...650L..11M,2018ApJ...863...78C,2020A&A...642L..19G,2022A&A...659A.155G}, and
the accurate measurement of proper motions is key to identify old globular clusters in the vicinity of the
Galactic center and to measure their orbital properties \citep{2021A&A...648A..86M,2023arXiv231216028M}.
\\

The \textit{VISTA Variables in the Via Lactea} \citep[VVV,][]{vvv2010} is a near-infrared survey of the
Galactic bulge and most of the Disk obtained with the wide-field camera mounted at the 
Visible and Infrared Survey Telescope for Astronomy (VISTA, located at the Paranal Observatory in Chile).
The VVV survey covers $\sim$\,528\,sq.\,deg of some of the most complex regions of the Milky Way in terms of high extinction
and crowding. Its extension, VVVx \citep{2018ASSP...51...63M}, re-observed the same regions covered by its predecessor, and observed for the first time 
new areas that were not included in the original VVV plan. The combination of VVV and VVVx data provides an incredible dataset
covering about 1700\,sq.\,deg observed between 20 and 300 times, with a temporal baseline of up to 10 years.

The VVV survey is carried out with the near-infrared, wide-field imager VIRCAM (VIsta InfraRed CAMera).
VIRCAM is an array of 16 Raytheon VIRGO Mercury Cadmium Telluride 2048$\times$2048-pixels
detectors, arranged in a 4$\times$4 grid.
The gap between the VIRCAM detectors are 42.5\% and 90\% of the detector's size along 
the $x$ and $y$ directions, respectively. This layout allows VIRCAM to cover a 
field of view of 0.6 deg$^2$ in a single pointing 
(a ``pawprint'', following the official nomenclature).
The VVV survey observed a given patch of sky (a ``tile'') with a 3$\times$2 mosaic,
covering about 1.4$\times$1.1 deg$^2$ without gaps.
The Cambridge Astronomical Survey Unit (CASU) is responsible for the data reduction,
catalog generation, and calibration of both photometry and astrometry
\citep{2004SPIE.5493..411I,2010ASPC..434...91L}.
The survey lasted five years, from 2010 to 2015, and observed the Disk and Bulge between 50 and 80
times in the $K_{\rm S}$ filter. Additionally, two epochs in $Z$, $Y$, $J$ and
$H$ filters were acquired at the beginning and at the end of the survey.
The VVV and VVVx surveys were mainly designed for photometric variability studies, but their multi-epoch strategy 
also enables astrometric analyses,
as demonstrated by, e.g., \cite{libra2015} and \cite{virac}.

\citeauthor{libra2015}\,(\citeyear{libra2015}, hereafter Paper\,I) presented a new pipeline to
process the VIRCAM data. They used a calibration field centered on the globular cluster
NGC\,5139 to derive a new geometric distortion (GD) solution for the VIRCAM detector,
which enables high-precision astrometry with the VVV data.
They reprocessed the data of the VVV tile containing the globular
cluster NGC\,6656 and derived proper motions with a precision of $\sim$1.4\,mas\,yr$^{-1}$,
using 45 epochs over a time baseline of 4 years.

\cite{virac} presented an astrometric catalog (\textit{VIRAC})
of proper motions and parallaxes for the entire VVV area
based on the CASU pipeline data;
this catalog contains 312 million sources with proper motions
and 6935 sources with parallaxes, and currently represents the largest
astrometric catalog derived from VVV. Their proper motions
achieve a precision of better than 1\,mas\,yr$^{-1}$ for bright
sources, and few mas\,yr$^{-1}$ at $K_{\rm S}=16$. However,
their astrometry is not absolute, as they did not have at
the time the {\it Gaia} all-sky reference frame to anchor their
positions. Moreover, uncertainties on the the relative to
absolute proper motion calibration limit the accuracy of
investigations on large Galactic scales.
\\

In this paper, we use the VVV data (not VVVx) to extend the {\it Gaia} astrometry to reddened sources in the Galactic plane.
In Fig.\,\ref{fig:source_d} we show the coverage of the VVV survey in Galactic coordinates, superimposed to a source-density map
of the {\it Gaia} Data Release 3 \citep[DR3][]{gaiadr3} catalog. We highlighted in light blue the two tiles considered
in this work, namely \texttt{b248} (South-East of the Galactic center)
and \texttt{b333} (which contains the Galactic center).
The density map shows that the number of sources accessible to {\it Gaia} in the Galactic plane is limited by dust and crowding.
As anticipated, this limit makes certain investigations unfeasible in this region, and it is the primary motivation
behind our work. 
We chose these two fields to test our methods in two environments with very different densities
of {\it Gaia} sources: the tile \texttt{b248} has about 350 {\it Gaia} sources per sq.\,arcmin, whereas the tile \texttt{b333}
has an average of 25 sources per sq.\,arcmin.
As in \citetalias{libra2015}, we 
use spatially variable empirical, effective point-spread functions \citep[ePSFs; see e.g.,][]{2006A&A...454.1029A} to precisely
measure positions and fluxes of all sources in any given VVV image, and adopt local transformations
\citep{2006A&A...454.1029A} to collate multiple single-image astro-photometric catalogs, to minimize systematic
errors such as residual GD and atmospheric effects, and to achieve the best astrometric
precision possible.
In addition, we significantly improve over previous efforts by: 
(i) employing a combination of first- and second-pass photometric stages specifically
designed to improve measurements in crowded fields and for faint stars, and (ii), linking our astrometry to that of the 
{\it Gaia} DR3 catalog, as done in, e.g., \cite{2018MNRAS.481.5339B,2020MNRAS.494.2068B} and \cite{2021MNRAS.500.3213L}. 

The paper is organized as follows. In Sec.\,\ref{sec:dr} we describe the data reduction process, the construction of the master
frame by leveraging on the {\it Gaia} DR3 catalog and the photometric registration. Section\,\ref{sec:ast} provides an overview
of the proper motions fit procedure. Section\,\ref{sec:gaia} shows the consistency of our catalog with respect to {\it Gaia}.
In Sec.\,\ref{sec:comp}, we compare our results with the current public release of the {\it VIRAC} catalog (version 1) and we show the improvements
enabled by our method. In Sec.\,\ref{sec:px}, we present an application of our new data reduction
by measuring the parallax of a sample of sources, and compare them with those in the {\it Gaia} catalog.
Finally, in Sec.\,\ref{sec:dra} we outline our data reduction strategy and
conclude the paper in Sec.\,\ref{sec:concl} with a summary of our work.

\section{Data reduction}
\label{sec:dr}

\begin{figure}
    \centering
    \includegraphics[width=\columnwidth]{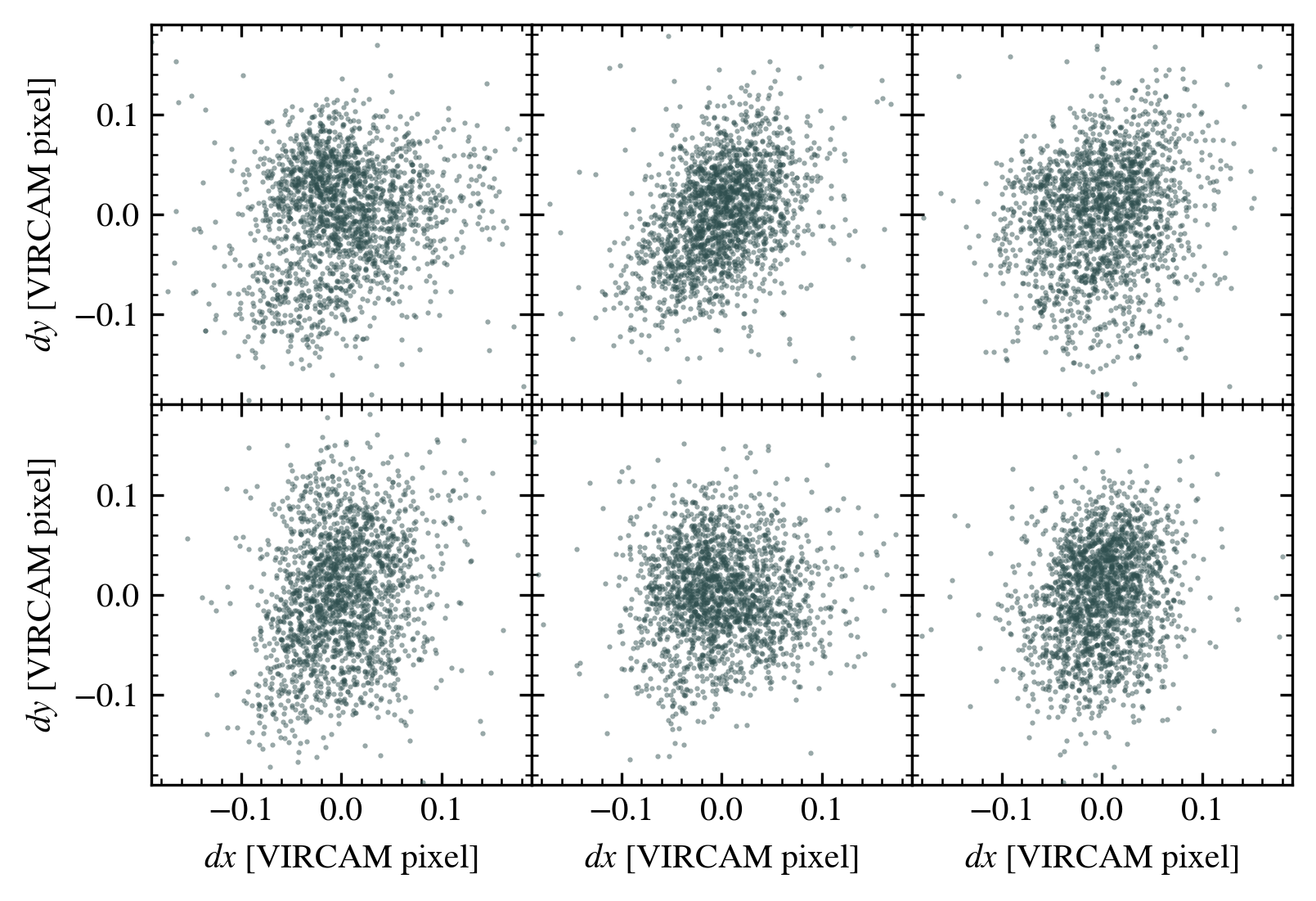}
    \caption{Positional residuals obtained by cross-matching the catalogs of a
    set of consecutive images with \textit{Gaia} after the GD solution 
    derived with the VVV dataset is applied. Each panel corresponds to a distinct image. 
    Notice that the residuals display different trends, preventing precision of less than $\sim$\,0.1 pixels
    ($\sim$\,35\,mas) in a single exposure with this approach.
    }
    \label{fig:imgmot}
\end{figure}

\begin{figure}
    \centering
    \includegraphics[width=\columnwidth]{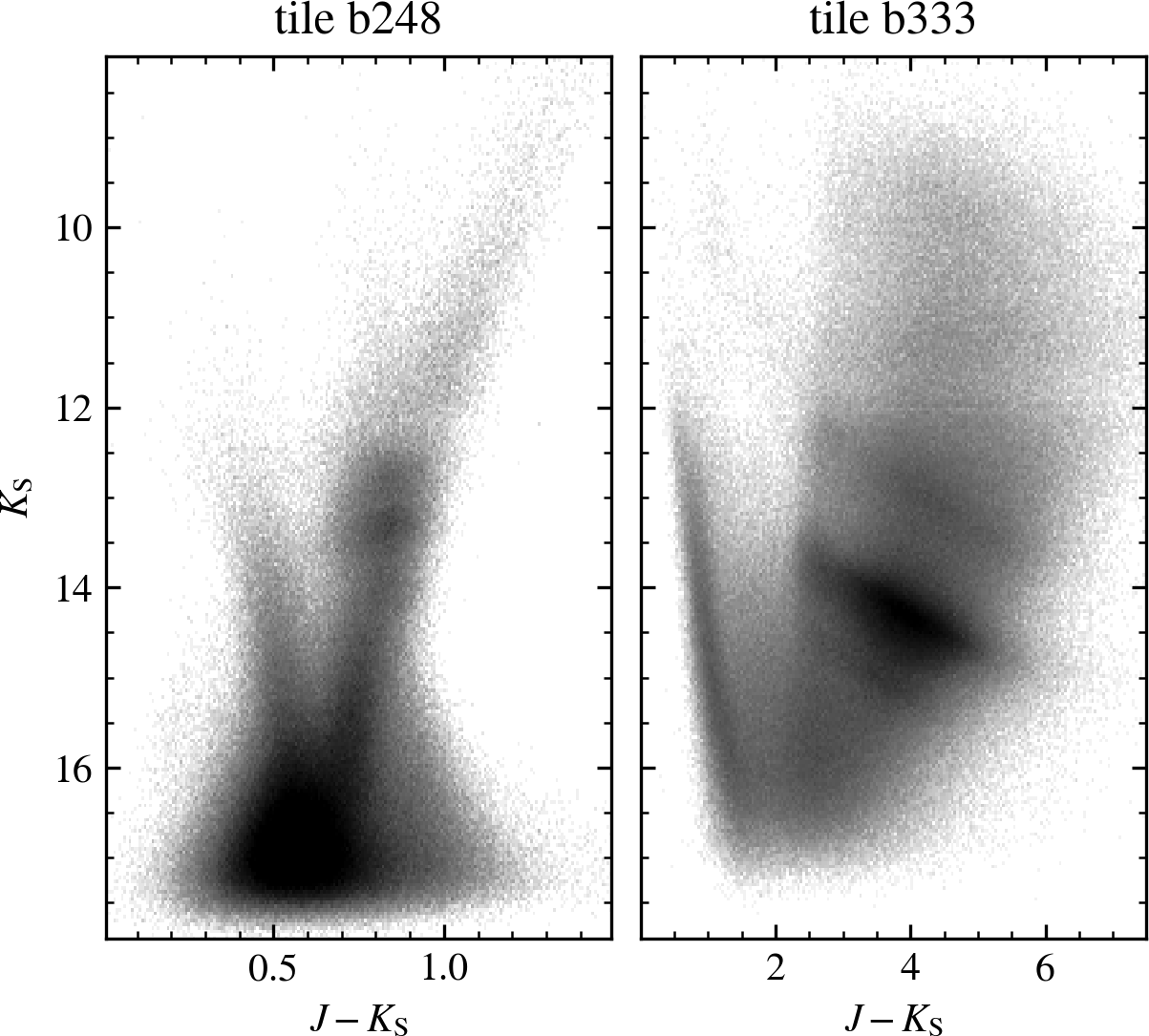}
    \caption{Color-magnitude diagrams for tiles \texttt{b248} (left) and \texttt{b333} (right).
    See the text for details.}
    \label{fig:cmds}
\end{figure}

In the first stage of the data reduction, i.e. ``preliminary photometry'', we focus on the bright, unsaturated members.
These sources are used to define the master frame by cross-matching
their positions with those measured by {\it Gaia}.
Preliminary photometry was obtained as described in
\citetalias{libra2015}. We started from the pre-reduced images downloaded
from the CASU archive and we treated each of the sixteen detectors separately.
Our pre-processing routine is responsible for applying
a series of cleaning algorithms to flag cosmic rays and mask bad
pixels and saturated stars.

For each detector, we derived a $5\times5$ array of effective point-spread
functions (ePSFs) using bright, isolated and unsaturated sources
as described by \cite{2006A&A...454.1029A}.
We used these ePSF models to measure positions and fluxes of the sources in each image.
Our preliminary photometric catalogs contain positions, instrumental magnitudes
defined as $-2.5\cdot\log(\rm flux)$,
and a parameter called quality-of-fit (\texttt{QFIT}), which represents the goodness of
the ePSF-model fit for each star 
\citep[see, e.g.,][]{2008AJ....135.2114A}. A \texttt{QFIT} value close to 0 represents
a good ePSF fit. In our VVV catalogs, bright and isolated sources typically have \texttt{QFIT}\,$<0.15$.
Bright, unsaturated stars close to saturation (with instrumental magnitudes between $\sim$\,$-13.5$ and $\sim$\,$-12$,
depending on the detector) show higher ($>0.2$) \texttt{QFIT} values. These sources
appear to exhibit inaccuracies in their flux measurements also in the {\it VIRAC}
catalog (see Fig.~\ref{fig:cmds_virac}).
What causes these bright stars, which should have a \texttt{QFIT} close to 0, to behave differently from the other bright, unsaturated sources is unclear. 
Nevertheless, we treated them as saturated and excluded the centermost pixels from the ePSF fitting to 
improve their photometry and astrometry.
The threshold used to identify these objects was empirically defined for each detector.

The calibration of the GD cannot be performed 
via auto-calibration techniques with the VVV data, as the dither pattern 
is not suited for this purpose 
\citep[see, e.g.,][]{2014A&A...563A..80L,2021MNRAS.503.1490H}. 
We attempted to derive the GD solution by exploiting the
\textit{Gaia} DR3 catalog, following the procedure described in
\cite{2022MNRAS.515.1841G,2023MNRAS.524..108G}.
However, given the short exposure time of the VVV images ($\sim$\,4\,s), we
were ultimately limited by the atmospheric image motion. 
In fact, the minimum exposure time needed to mitigate the
impact of large-scale semi-periodic and correlated atmospheric noise, which can
adversely affect ground-based astrometry, is approximately 30 seconds, 
as determined by, e.g., \citet{2002AJ....124..601P,2006PASP..118..107P} or \citet{2014A&A...563A..80L}.
To show the effect of image motion, we cross identified the stars
in the astro-photometric catalogs of a set of consecutive VVV images and in the Gaia DR3 catalog.
Gaia positions were propagated at the epoch of the VVV images by means of the {\it Gaia} DR3 proper motions and projected on the tangent plane of each VVV exposure.
We then used six-parameter linear transformations to transform the GD-corrected positions in each VVV frame on to the {\it Gaia} reference frame system.
Figure\,\ref{fig:imgmot} shows the positional residuals generated by this
comparison for a set of consecutive VVV images. Even though these images were 
taken within less than one minute, the residuals show different trends.
Computing the GD solution from these images results into a null
mean correction, as the random residuals due to image motion
cancel out. For this reason, it is not possible to improve
the GD solution derived in \citetalias{libra2015} using the
VVV data itself.

We then corrected the VVV raw positions
with the GD solution of \citetalias{libra2015}, which we consider more reliable as it was
computed from well-exposed images of a calibration field specifically observed to calibrate the GD.
We will explain later, in Sec.\,\ref{sec:bore}, how we mitigated the effect of image motion.

Besides the GD, we also needed to consider projection effects resulting from the
large dithers of the images and the wide field of view of VIRCAM \citepalias[see discussion in][]{libra2015}.
As a result, images lie in different tangent planes, and we need to account for this as done in, e.g.,
\cite{2022MNRAS.515.1841G}.
To do so, after correcting the raw positions with the GD solution,
we used the information contained in the \texttt{fits} header of each
image to project the detector-based coordinates onto equatorial coordinates,
as in \cite{2018MNRAS.481.5339B},
adopting a pixel scale of 0.339 arcsec/px \citepalias{libra2015}.
We then projected all positions from all catalogs back into a common plane, using as tangent
point the average pointing position of all the images of the same tile.
After this procedure, all single-exposure catalogs lie on the same tangent plane.

\subsection{The master frame}

A key step for our goal is the construction of a common reference frame,
hereafter ``master frame'', where we can combine our images.
Thanks to the {\it Gaia} mission, we already have an all-sky, absolute reference
frame to which we can anchor our astrometry.
We used the \textit{Gaia} DR3 catalog to define the master frame
as follows:
\begin{itemize}
    \item We propagated the \textit{Gaia} positions to the corresponding epoch of each VVV pawprint set
    using {\it Gaia} proper motions.
    \item We projected these adjusted positions onto the same tangent plane adopted for the tile.
    \item We cross-matched the sources in each single-exposure catalog 
    with those in the \textit{Gaia} DR3 catalog, and derived the coefficients of 
    the six-parameters linear transformations that bring the image-based coordinates 
    onto the \textit{Gaia} absolute reference frame.
\end{itemize}

\subsection{Photometric registration}

We registered the photometry to the
\textit{2MASS} photometric system \citep{2006AJ....131.1163S}
using the \textit{Gaia}-\textit{2MASS}
cross-matched table available in the
\textit{Gaia} archive. We selected the best measured sources in both
our catalogs and \textit{2MASS}, and use these sources to calculate 
the photometric zero-points to transform the instrumental magnitudes
measured in each single image into the \textit{2MASS} photometric system. 

The \textit{2MASS} filter passbands
are slightly different from those of the VISTA filters, and a more precise calibration
would need to account for several factors, as described in \cite{2018MNRAS.474.5459G}
and \cite{2020ExA....49..217H}.
However, since in this work we are focused on obtaining high-precision astrometry,
we neglected second-order corrections. 
In fact, the largest term in the photometric-calibration equation between
the \textit{2MASS} and VISTA magnitudes is the color term,
and it is of the order of few hundredths of magnitudes at most.
As such, the correction introduced by this term would be very small.
Nonetheless, we plan to perform a more accurate calibration in the next releases,
including airmass, extinction and color terms.

\subsection{Second pass photometry}
\label{sec:2pass}

The ``second-pass'' photometry has been performed using a version of the 
software \texttt{KS2} 
\citep[an evolution of the code presented in][developed for \textit{HST}]{2008AJ....135.2055A}, 
opportunely modified to work with VIRCAM images, and adapted to wide-field imagers
by \cite{2022MNRAS.515.1841G,2023MNRAS.524..108G}.
The software is designed to obtain deep photometry in crowded fields,
by iterating a find-measure-subtract routine, that employs all images
simultaneously to improve the finding of faint sources.
A description of \texttt{KS2} can be found in, e.g., \cite{2017ApJ...842....6B} and \cite{2021MNRAS.505.3549S}.
Briefly, the flux is measured by fitting the ePSF
to the inner $5\times5$ pixels of the source after subtracting the local sky,
using the appropriate ePSF for each image, and then averaged out between all
the exposures.
Stars measured in the previous iteration are subtracted 
from the image at each step. 
The second-pass photometry has been carried out separately for each epoch in the $K_{\rm S}$ filter:
we considered as an epoch each set of images taken in the same day.
A list of stars (derived from the preliminary photometry) is given as input
to the routine, and it is used to construct a weighted mask around bright sources
which helps to avoid PSF-related artifacts.
In addition to the averaged master frame positions, the \texttt{KS2} software outputs a file that contains, 
for each source, the raw position and neighbor-subtracted flux as measured in every single exposure.
In addition to positions and fluxes, \texttt{KS2}
also outputs a few diagnostic parameters 
\citep[see, e.g.,][]{2009ApJ...697..965B},
that can be used to reject poorly measured sources and
detector's artifacts, or to identify galaxies.
In addition to the $K_{\rm S}$ exposures, we also performed the second-pass photometry
on all the $J$ images of each tile, that we used {\it only} to build the color-magnitude diagrams.
The color-magnitude diagrams of the two tiles analyzed in this work are shown in
Fig.\,\ref{fig:cmds}. The non-physical drop in the number of sources around $K_{\rm S}$\,$\sim$\,12 is due to
our quality cut in the \texttt{QFIT} parameter:
at this magnitude level there is a transition between unsaturated and saturated sources, 
and it is here where most of the stars showing unexpectedly-high \texttt{QFIT} values lie, as discussed in
Sect.~\ref{sec:dr}. Some of these problematic sources end up being measured as unsaturated
as the thresholds that we set to identify these objects are not perfect,
resulting in an overall high \texttt{QFIT} value.

\subsection{The boresight correction}
\label{sec:bore}
\begin{figure}
    \centering
    \includegraphics[width=\columnwidth]{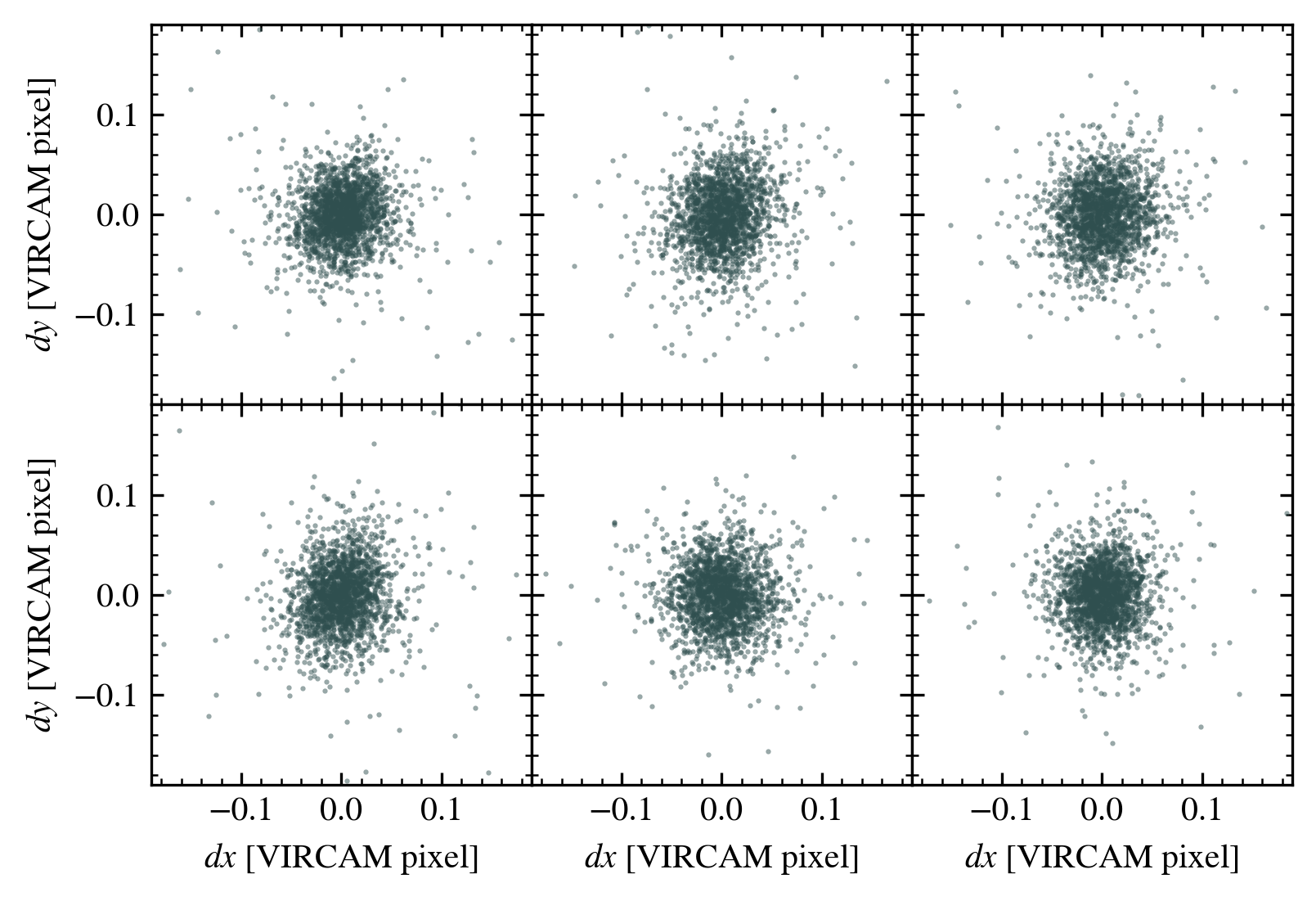}
    \caption{Similar to Fig.\,\ref{fig:imgmot}, but after the boresight correction is applied.
    The dispersion of the residuals is about 0.035 pixels ($\sim$\,12\,mas) along each coordinate.}
    \label{fig:bore}
\end{figure}

Image motion poses a severe limitation to 
the astrometric precision achievable with VVV data. Indeed,
it leads to local systematic position errors whose pattern
changes even between subsequent exposures, up to $\sim$\,0.15\,pixels 
($\sim$\,50\,mas, see Fig.\,\ref{fig:imgmot}).
This is far more than the single-measurement astrometric precision 
enabled by the GD solution,
which has been shown to reach $\sim$\,8\,mas \citepalias{libra2015}.
We employed a local mitigation to the effects of image motion on a given source (the target source)
through a so-called ``boresight'' correction \citep[see, e.g.][]{2010ApJ...710.1032A,2014MNRAS.439..354B}. 
Our correction leverages on the \textit{Gaia} catalog,
as it provides a sufficient number of reference \textit{Gaia} sources even in
regions towards the Galactic center.
The number of neighboring reference sources for the correction
is a compromise between the need for high statistics and for the correction
to be as local as possible, even in regions where the Gaia source density is low.
We achieved this by requiring at least 15 reference sources within a circular region of radius at most 300 pixels (and at least 50 pixels) from the target source.

For each epoch, we calculated the mean position of each source as
follows:
\begin{itemize}
    \item we applied the GD correction to the raw positions of the sources
    in each image;
    \item we projected the corrected positions onto the sky, and then we
    project them back onto the common tangent plane of the tile;
    \item using well-measured sources in common with \textit{Gaia} (excluding the target),
    we computed the transformations to bring the positions 
    from the tangent plane to the master frame;
    \item for each target star, we selected the neighboring sources in common with \textit{Gaia}
    in each image, and
    we calculated the residuals between their
    positions transformed into the master frame
    and those given by \textit{Gaia} (again, excluding the target star).
    The mean residual gives the boresight correction for each target source of each image.
    \item We then transformed the positions of the target star as measured
    in the single images into the master frame, and apply to these values
    their boresight correction.
    \item Finally, we calculated the average position, to which we associate
    the error on the mean as an estimate of the uncertainty.
\end{itemize}

Figure\,\ref{fig:bore} is similar to Fig.\,\ref{fig:imgmot}, but with the positional residuals computed
after the image-motion correction is applied. It is clear that 
the distribution of the residuals is much tighter than before, with a dispersion of about 12\,mas
in each coordinate in a single exposure, a value that is
compatible with the results obtained in \citetalias{libra2015}.

\section{Proper motions}
\label{sec:ast}

\begin{figure}
    \centering
    \begin{subfigure}[t]{\columnwidth}
        \centering
        \includegraphics[width=\textwidth]{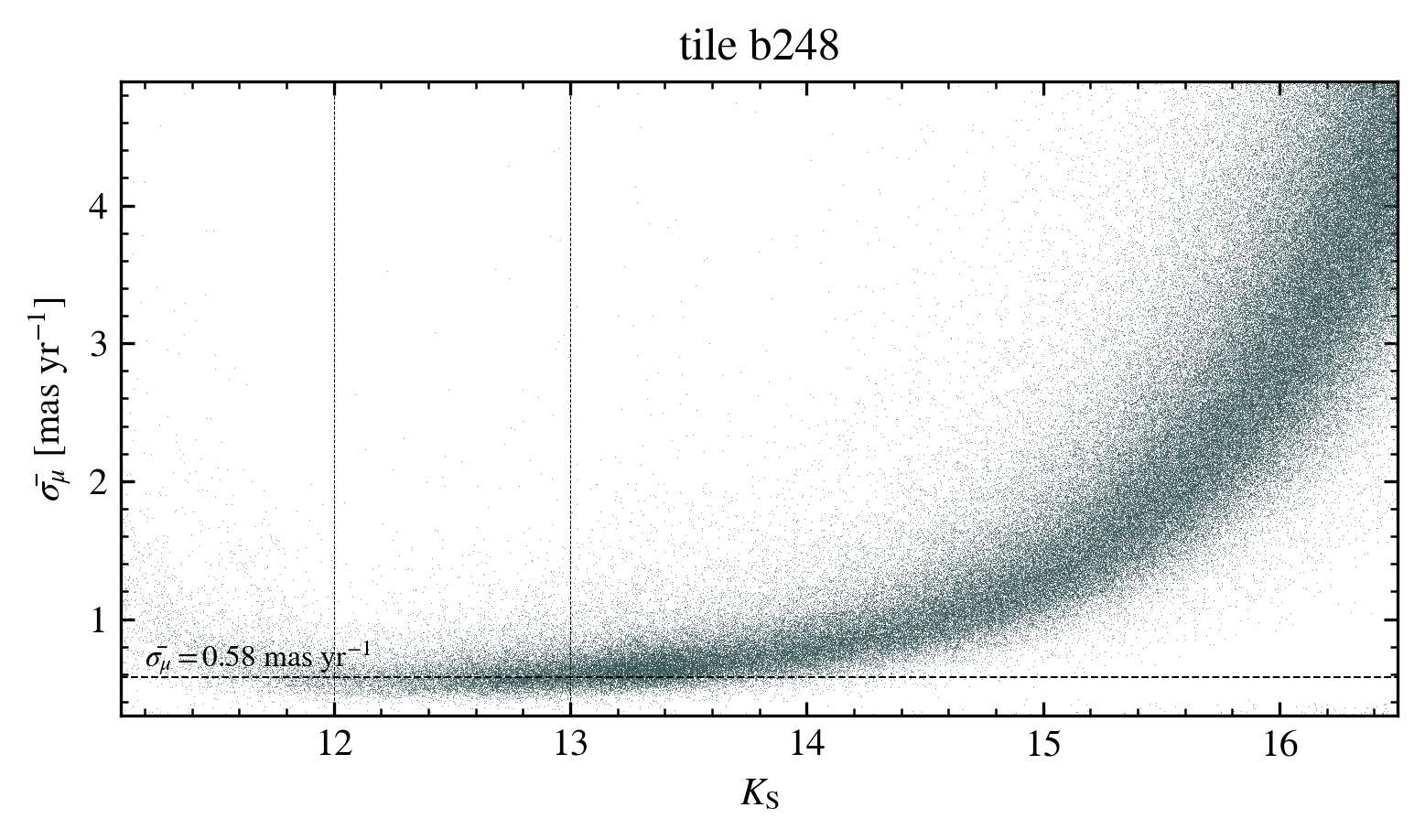}
    \end{subfigure}%
    \\ 
    \centering
    \begin{subfigure}[t]{\columnwidth}
        \centering
        \includegraphics[width=\textwidth]{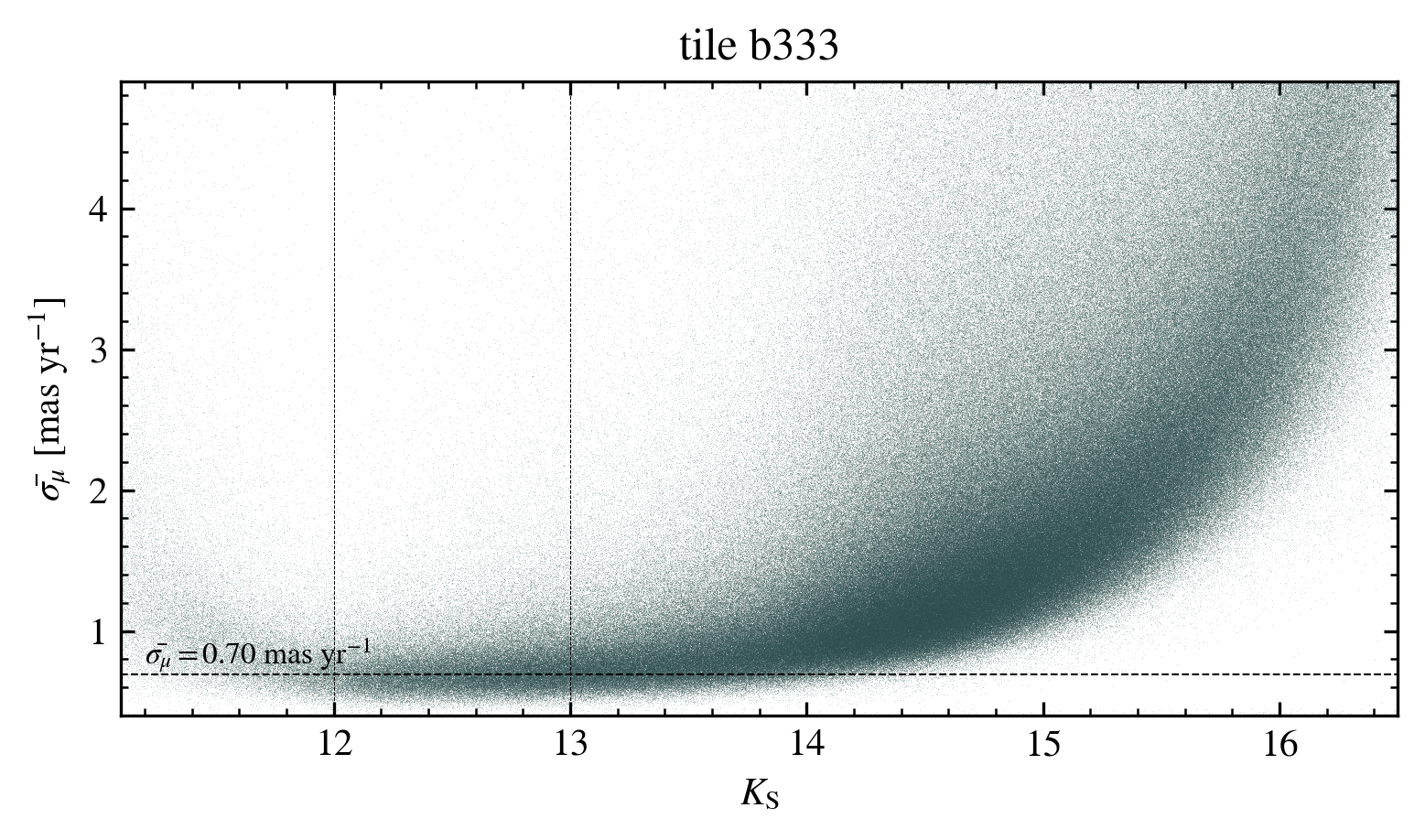}
    \end{subfigure}%
    \caption{Mean proper-motion errors
    for sources in tile \texttt{b248} (top)
    and tile \texttt{b333} (bottom) as a function of the $K_{\rm S}$ magnitude.
    The horizontal line represents the median error of the sources in the range
    $12<K_{\rm S}<13$. See the text for details.}
    \label{fig:pms_err}
\end{figure}

\begin{figure*}
    \centering
    \begin{subfigure}[t]{\columnwidth}
        \centering
        \includegraphics[width=\textwidth]{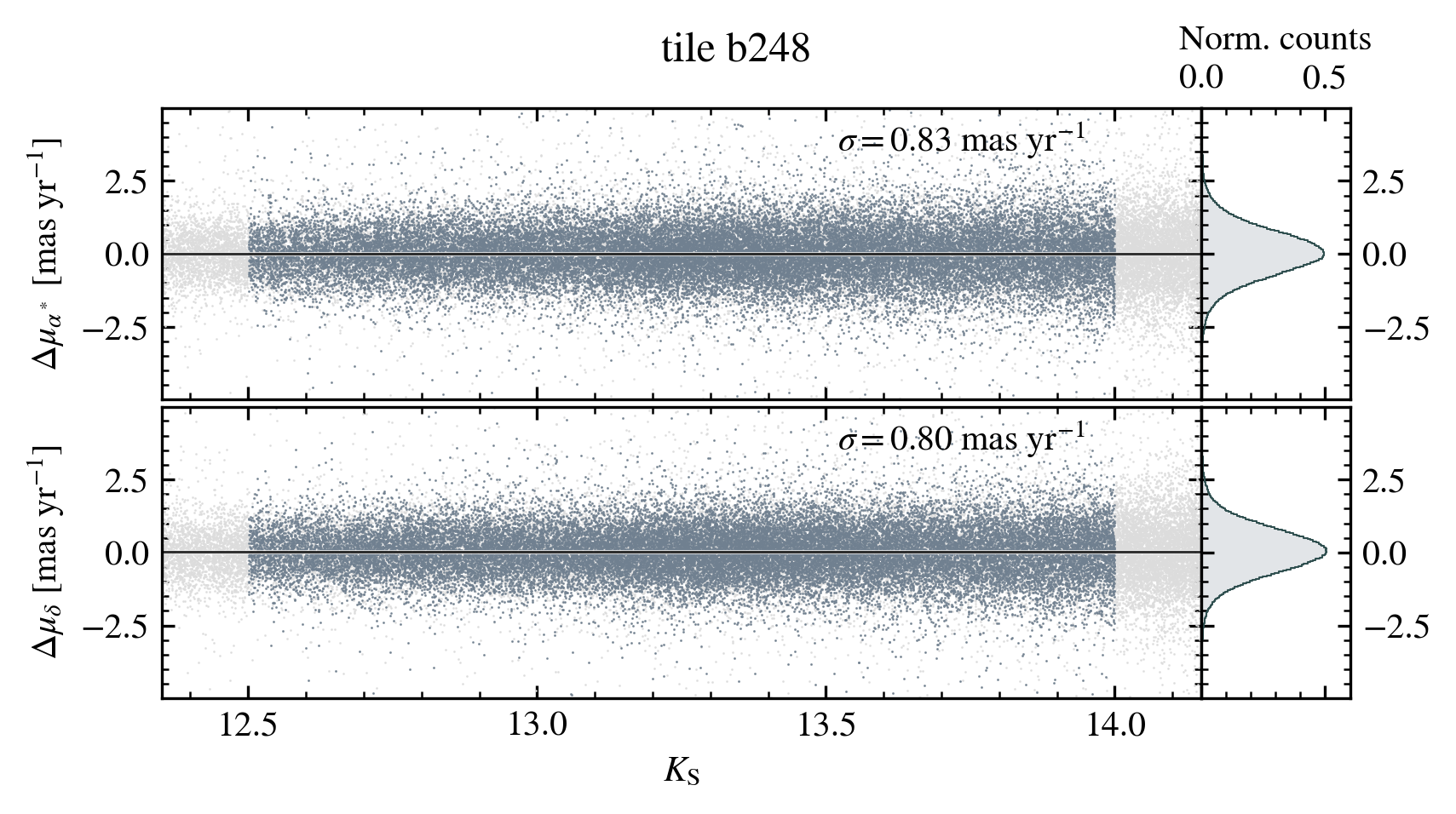}
    \end{subfigure}%
    ~
    \centering
    \begin{subfigure}[t]{\columnwidth}
        \centering
        \includegraphics[width=\textwidth]{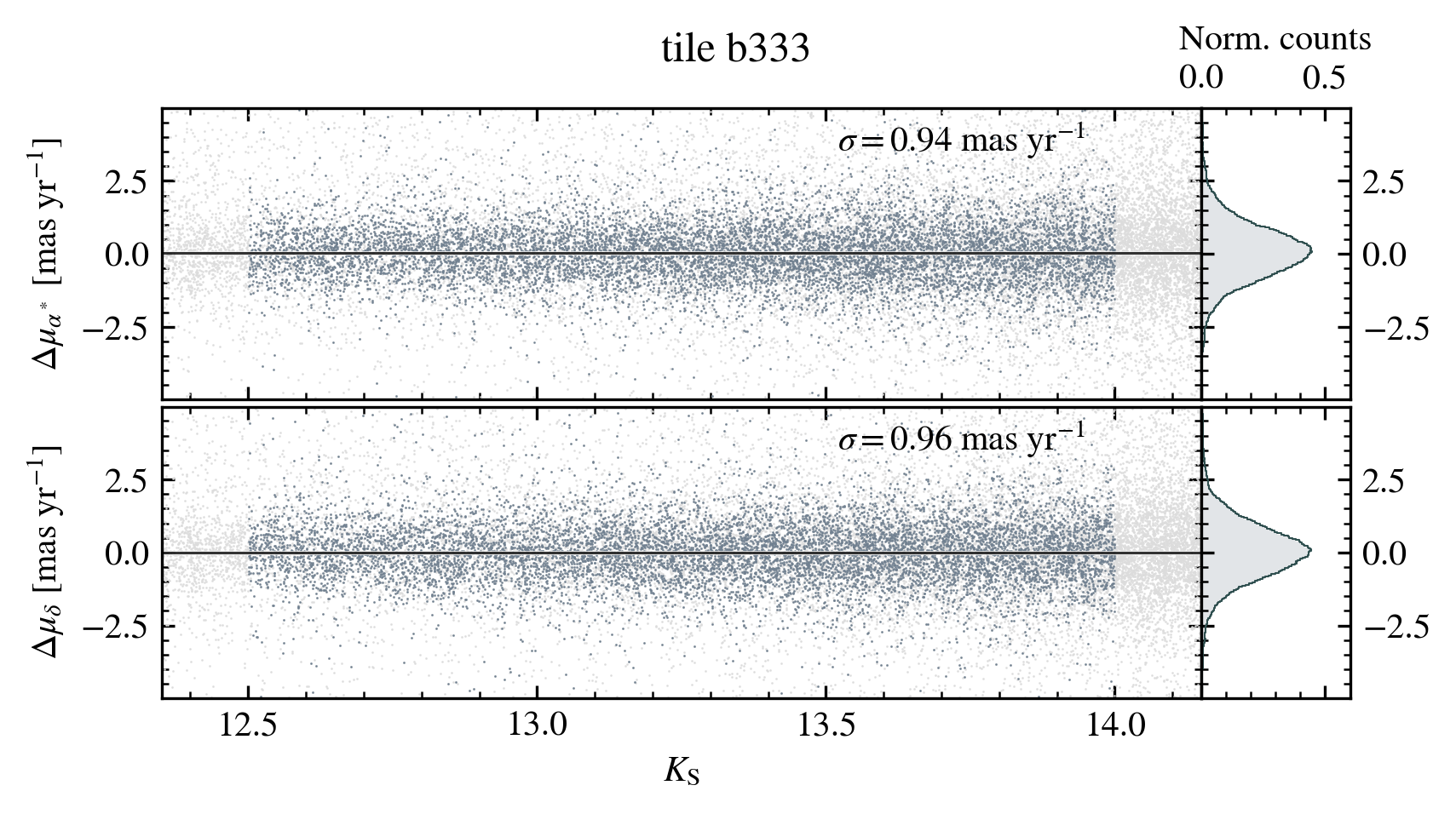}
    \end{subfigure}%
    \caption{Comparison between proper motions computed in this work and
        \textit{Gaia} proper motions for tile \texttt{b248} (left) and
        tile \texttt{b333} (right).
        The black line is the median offset,
        calculated using the sources in dark grey.}
    \label{fig:pms_comp}
\end{figure*}

Proper motions were obtained via a maximum likelihood approach
using the affine-invariant Markov Chain Monte Carlo method
\texttt{emcee} \citep{2013PASP..125..306F}, to sample the parameter space.
This approach allows us to obtain the posterior probability distribution functions
for the quantities $\mu_x,\mu_y,x_0,y_0$, where $\mu_x,\mu_y$ are the displacements in the
$x$ and $y$ directions, and $x_0,y_0$ are the positions at the reference epoch $t_0$,
that we set to be $t_0=2013.0$.
We ran the Markov Chain Monte Carlo with 32 walkers, performing 5000 steps, with 200 burn-in steps,
allowing some scaling on the positional errors as an additional free parameter to be fitted.
The medians of the probability distribution functions give our final estimate of $\mu_x,\mu_y,x_0,y_0$,
and the errors on these quantities were computed as the average between the
$16^{\rm th}$ and $84^{\rm th}$ percentiles of the samples in the marginalized distributions.
The displacements in pixel\,yr$^{-1}$ are converted in 
$\mu_{\alpha^*},\mu_\delta$ by multiplying by the pixel scale that we adopted
(0.339\,arcsec pixel$^{-1}$), as the master frame axes are already oriented as North and East.
At odd with the {\it VIRAC} proper motions, which are relative, our proper motions are naturally
defined on an absolute system, since their computation is based on the {\it Gaia} reference frame.

Figure\,\ref{fig:pms_err} presents the mean proper-motion 
errors $\bar{\sigma_\mu}=(\sigma_{\mu_{\alpha^*}}+\sigma_{\mu_\delta})/2$
as function of the $K_{\rm S}$ magnitude.
Given the extreme crowding environment of the Galactic center,
proper-motion errors of tile \texttt{b333} are larger than those
of tile \texttt{b248}: the median proper-motion error
of sources in the range $12<K_{\rm S}<13$ -- i.e. best measured sources --
is 0.58\,mas\,yr$^{-1}$ in tile
\texttt{b248}, compared to 0.70\,mas\,yr$^{-1}$ in tile \texttt{b333}.
Sources with $K_{\rm S}\lesssim12$ are close to saturation/non-linearity regime,
and thus their proper-motion errors tend to increase.

\section{An extension of \textit{Gaia} into the Galactic plane}
\label{sec:gaia}

\begin{figure*}
    \centering
    \begin{subfigure}[t]{\columnwidth}
        \centering
        \includegraphics[width=\textwidth]{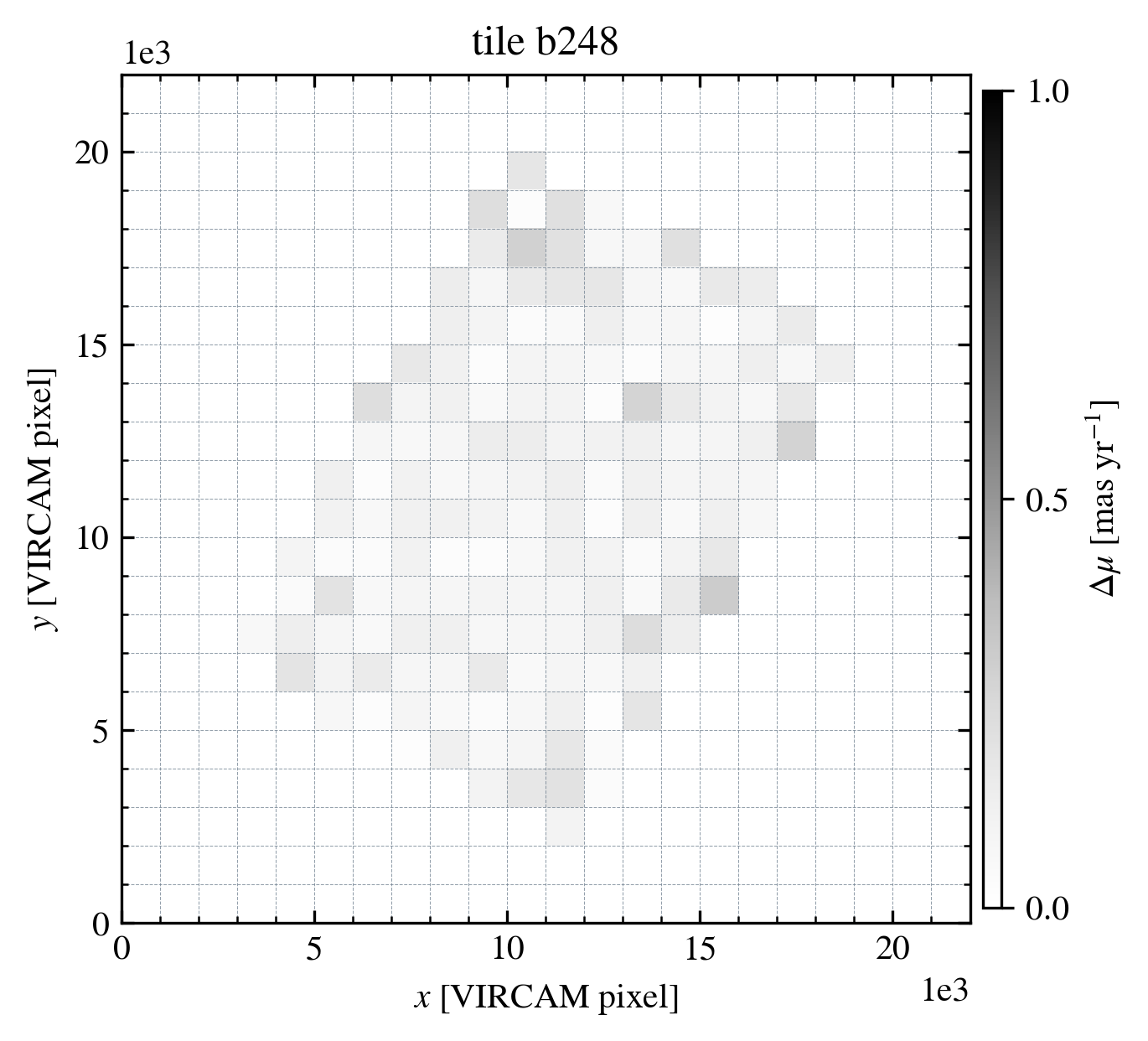}
    \end{subfigure}%
    ~
    \centering
    \begin{subfigure}[t]{\columnwidth}
        \centering
        \includegraphics[width=\textwidth]{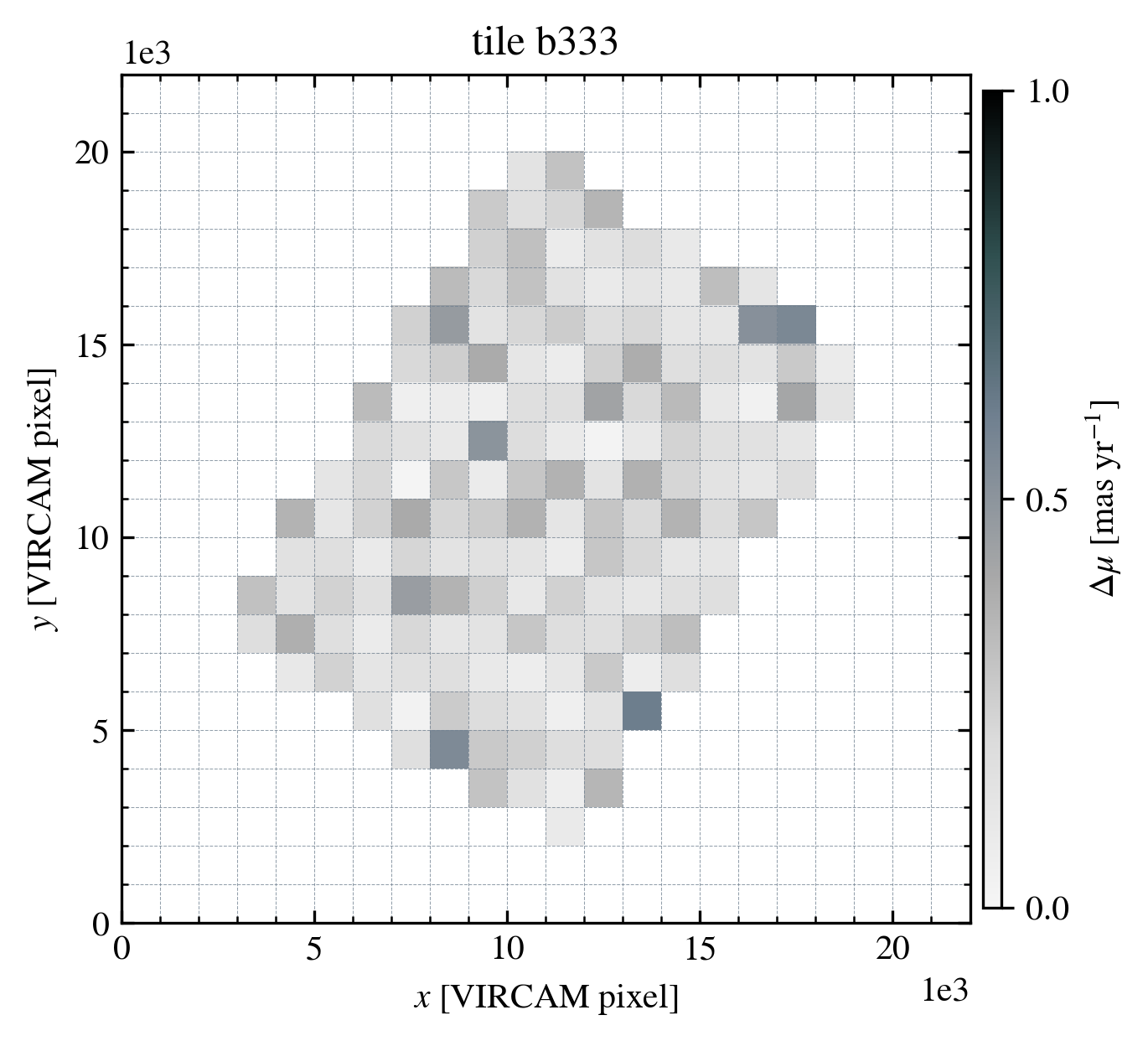}
    \end{subfigure}%
    \caption{Similar to Fig.~\ref{fig:pms_comp}, but as a function of position for tile \texttt{b248} (left) and
        tile \texttt{b333} (right). Each region is colored according to the $3\sigma$-clipped median value of the absolute deviation between \textit{Gaia}'s and our proper motions, according to the color bars on the right of each panel.}
    \label{fig:pms_comp_color}
\end{figure*}

\begin{figure}
    \centering
    \includegraphics[width=.95\columnwidth]{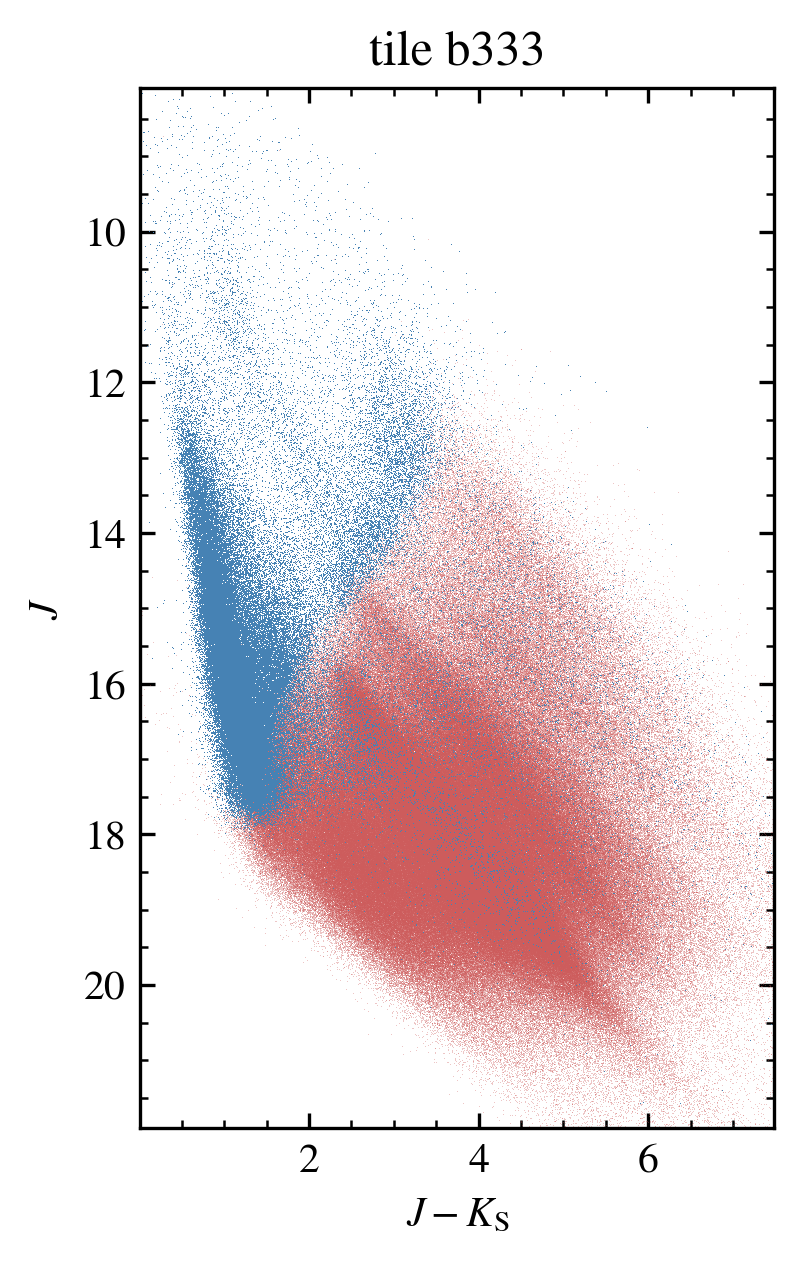}
    \caption{The $J$ versus $(J-K_{\rm S})$ color-magnitude diagram of the sources in tile \texttt{b333};
    blue sources are those also present in the {\it Gaia} catalog.}
    \label{fig:cmd_gaia}
\end{figure}

\begin{figure}
    \centering
    \includegraphics[width=\columnwidth]{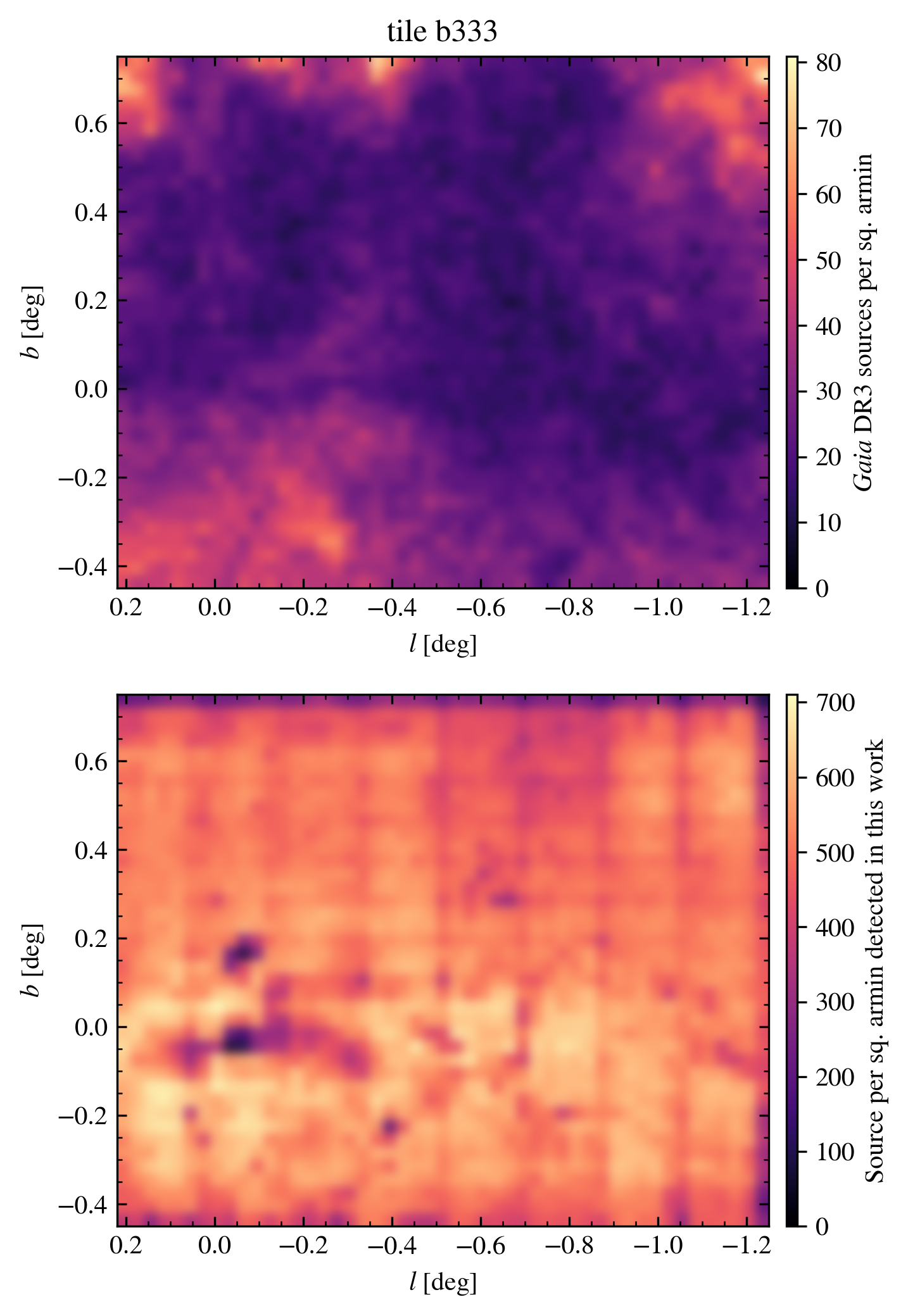}
    \caption{Source density of the {\it Gaia} DR3 catalog (top) and of our catalog (bottom) for tile \texttt{b333} in Galactic coordinates. See the text for details.}
    \label{fig:sdb333}
\end{figure}

Our astrometry is, by construction, linked to the {\it Gaia} absolute reference frame. We can
see the consistency between {\it Gaia}'s and our proper motions in Fig.\,\ref{fig:pms_comp},
where we show the residuals
between the two datasets,
$\Delta\mu_{\alpha^*},\Delta\mu_{\delta}$, for tile \texttt{b248} (left) and tile \texttt{b333} (right).
The normalized histograms of the residuals are shown in the right panel of the plots.
Dark gray points are the sources with proper-motion errors smaller than
2\,mas\,yr$^{-1}$, $K_{\rm S}$ magnitude between 12.5 and 14, {\it Gaia} $G$ magnitude between 13 and 18,
and measured in at least four individual images\footnote{This restriction excludes sources near the tiles' borders that, as a result of the VVV dither pattern, are covered by less than four exposures.
In future releases, we plan to use the data of adjacent tiles 
to increase the depth of coverage and measure proper motions in these regions.}.
These objects represent well-measured sources in both sets.
The horizontal black line represents
the median residuals calculated using the dark grey sources.
The standard deviation of the residual distributions are reported on the
top right corner of each plot. Given the essentially negligible errors
of the {\it Gaia} proper motions with respect to ours, the dispersion
can be attributed to the random errors on our astrometry.
In this regards, we notice that the dispersion is slightly larger than the
proper-motions errors obtained by our fit as shown in Fig.\,\ref{fig:pms_err}
for sources with $K_{\rm S}<14$, suggesting a possible underestimation of the proper-motion errors.

In Fig.\,\ref{fig:pms_comp_color} we show the local deviations between {\it Gaia}'s and our proper motions.
We divided the field in $\sim$1000$\times$1000 pixel regions and, for
each region, we calculated the 3$\sigma$-clipped medians ($\bar{\Delta\mu_{\alpha^*}},\bar{\Delta\mu_{\delta}}$)
of the proper-motions residuals.
We then computed the quantity $\bar{\Delta\mu}^2=\bar{\Delta\mu_{\alpha^*}}^2+\bar{\Delta\mu_{\delta}}^2$
using only well-measured sources, i.e.
dark grey points of Fig.\,\ref{fig:pms_comp}. From the left panel of Fig.\,\ref{fig:pms_comp_color}, we can notice 
that the proper motions of tile \texttt{b248} 
are in very good agreement with {\it Gaia}, 
and, as expected, the $\bar{\Delta\mu}$ distribution across the tile is flat, with negligible local systematic
deviations.
The right panel for the tile \texttt{b333} instead shows hints of some spatially-dependent systematic errors.
We verified that these systematics correlate
with the density of {\it Gaia} sources in each region,
with larger deviations corresponding to regions with very low density.
This correlation is expected as our technique relies on a local net
of {\it Gaia} sources to compute the boresight correction.
Except for these regions with fewer {\it Gaia} sources,
our proper motions for the tile \texttt{b333} are in good agreement with those in {\it Gaia},
with slightly larger residuals with respect to those exhibited by the
tile \texttt{b248}, given smaller {\it Gaia} source density in tile \texttt{b333}.

Our independent reduction of the VVV fields allows us to extend the {\it Gaia} astrometry to the
dense and obscured regions of the Galactic plane, providing significantly more new sources with positions and proper motions than what is available from the \textit{Gaia}\,DR3 catalog in the same region. Figure\,\ref{fig:cmd_gaia} shows a $J$ versus $(J-K_{\rm S})$ color-magnitude diagram of the sources in tile \texttt{b333}. Blue points represent stars in common with the {\it Gaia} DR3 catalog. All other objects are in red. It is clear that a large number of sources are not present in the {\it Gaia} catalog. Our catalog of tile \texttt{b333} contains more than 2\,millions sources with
proper motions and only $\sim$\,10\% of them are present in {\it Gaia} DR3.
In Fig.\,\ref{fig:sdb333} we show, for the same tile, the source density of the {\it Gaia} DR3 catalog (top) and of our catalog (bottom). 
We used all the sources in our catalog and all the sources in the {\it Gaia} catalog
for this plot. We computed the values binning the data
into $\sim$\,3\,sq.\,arcmin bins and smoothing out with a Gaussian kernel.
In the Galactic center we detect on average $\sim$\,20 times more sources
than {\it Gaia}, with regions where this values goes up to $\sim$\,60.

\section{Comparison with \textit{VIRAC}}
\label{sec:comp}

\begin{figure}
    \centering
    \includegraphics[width=\columnwidth]{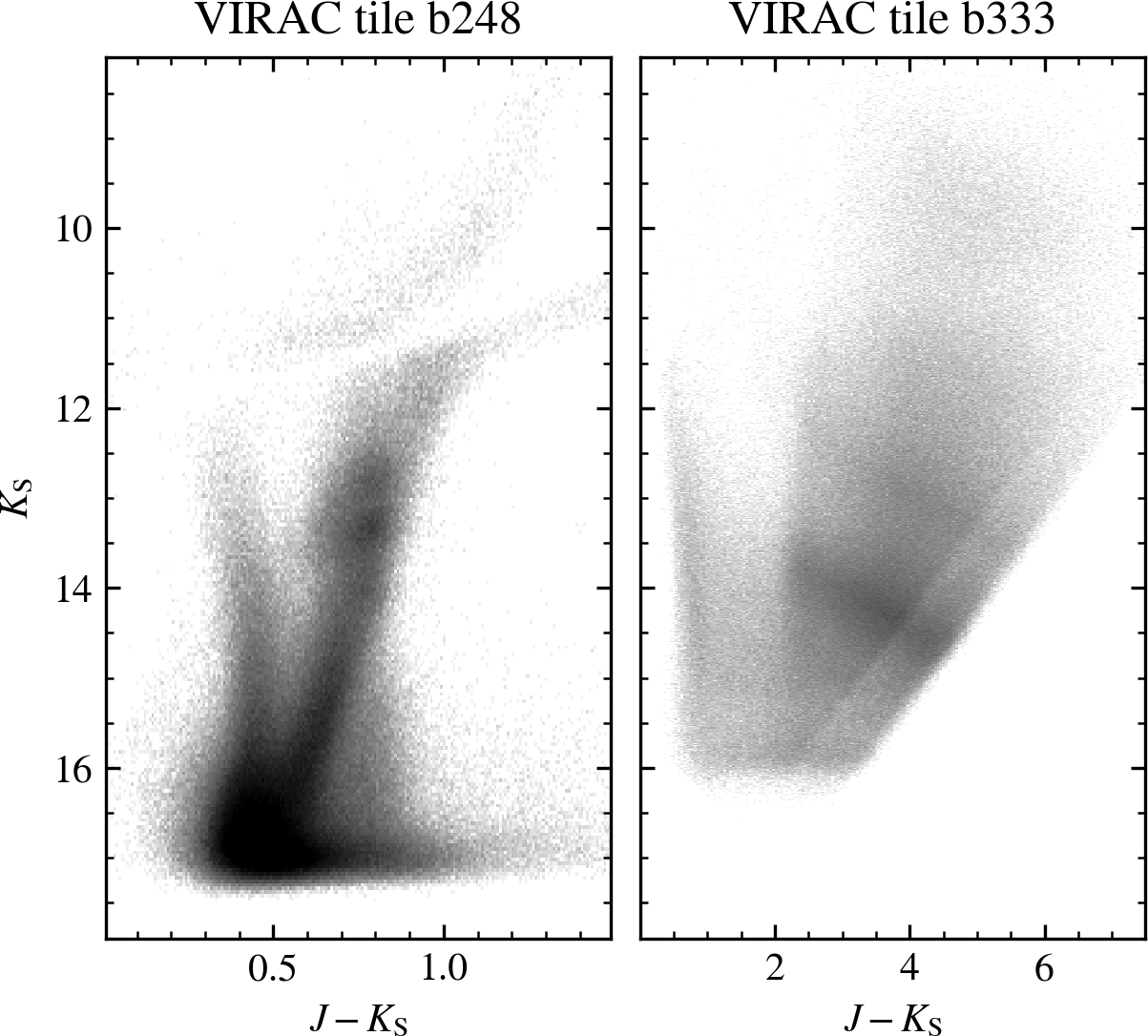}
    \caption{Similar to Fig.\,\ref{fig:cmds}, but for {\it VIRAC} sources.}
    \label{fig:cmds_virac}
\end{figure}

\begin{figure*}
    \centering
    \begin{subfigure}[t]{\columnwidth}
        \centering
        \includegraphics[width=\textwidth]{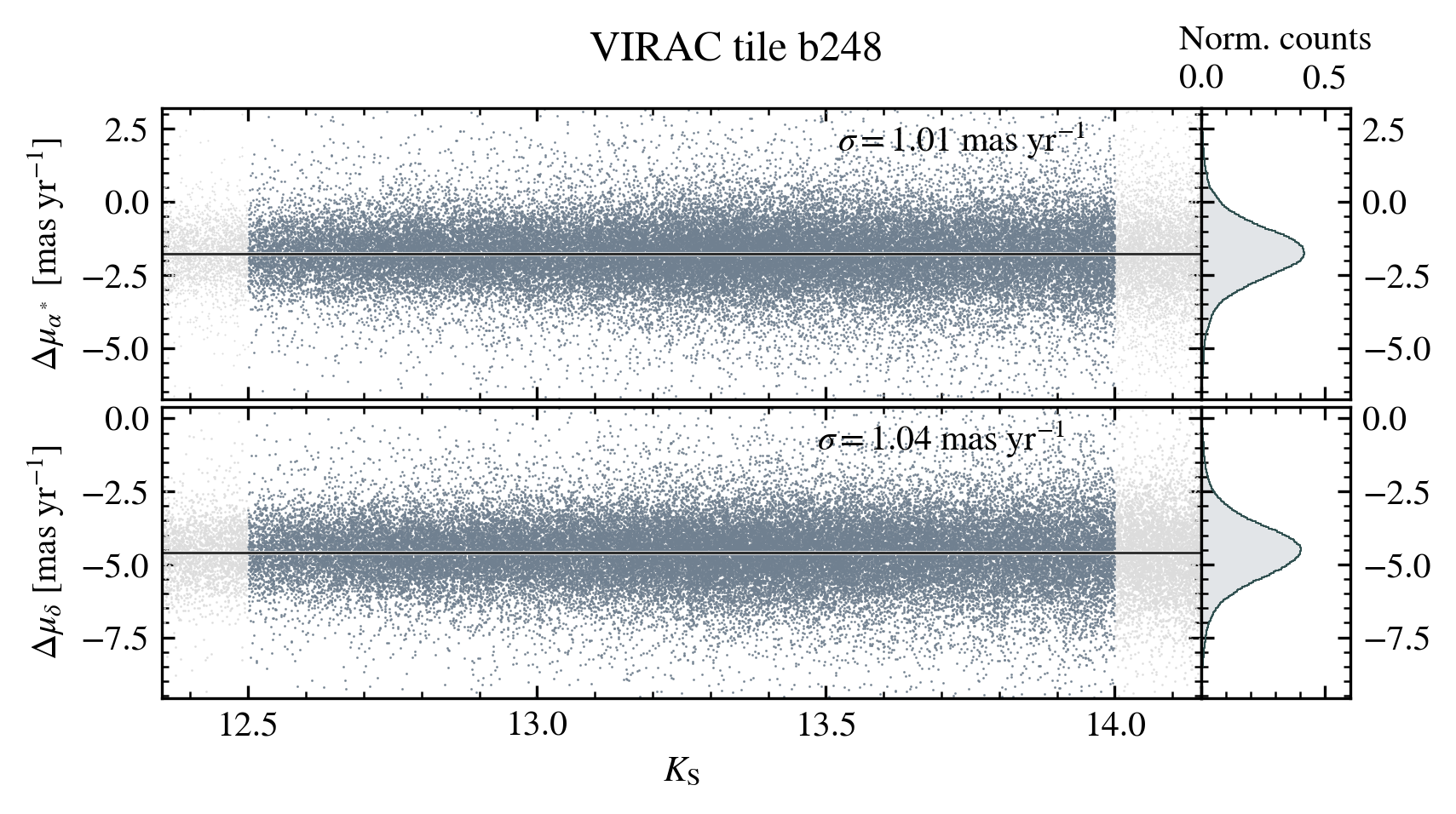}
    \end{subfigure}%
    ~
    \centering
    \begin{subfigure}[t]{\columnwidth}
        \centering
        \includegraphics[width=\textwidth]{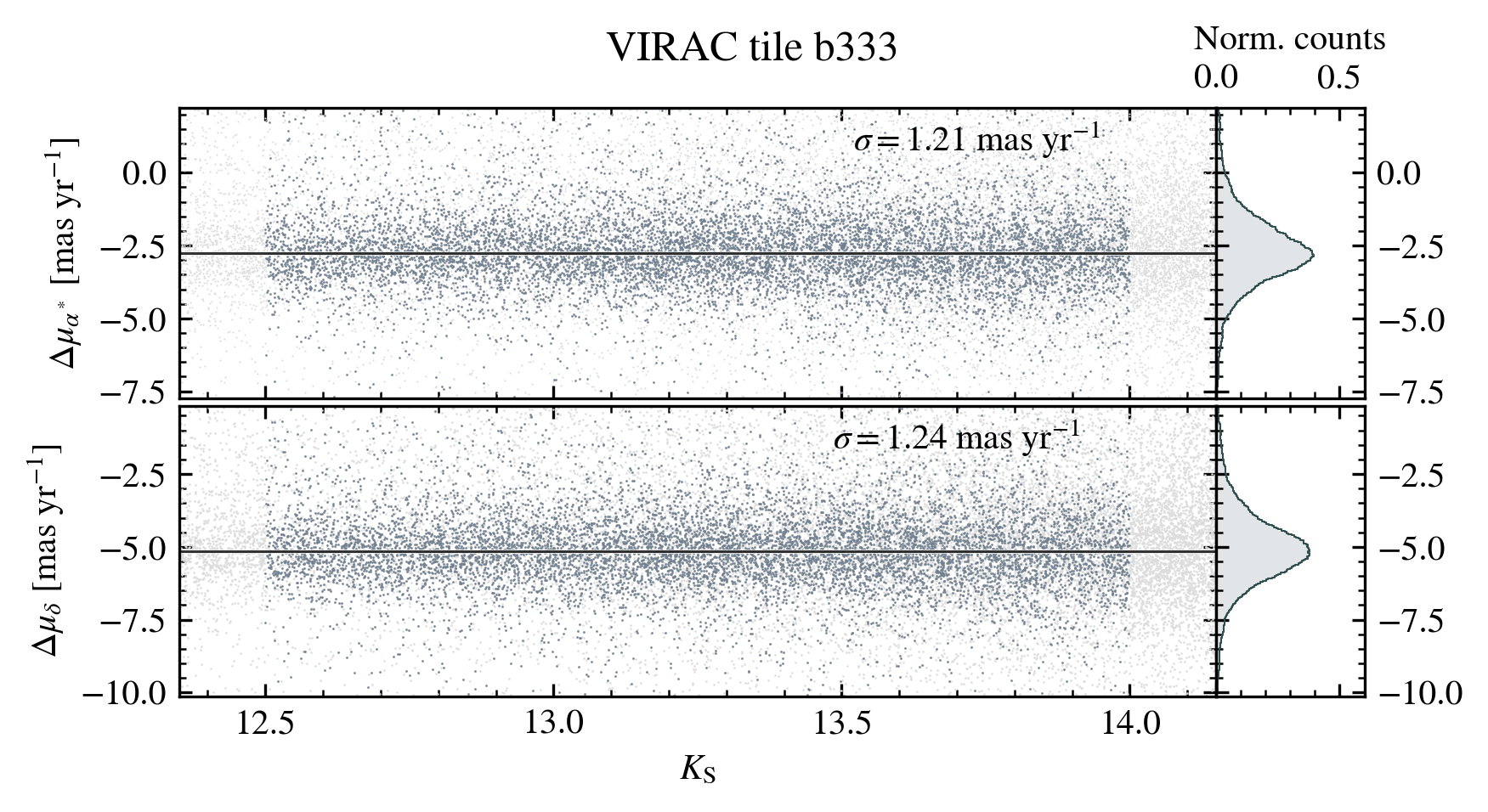}
    \end{subfigure}%
    \caption{Similar to Fig.\,\ref{fig:pms_comp}, but for {\it VIRAC} sources.}
    \label{fig:pms_comp_virac}
\end{figure*}

\begin{figure*}
    \centering
    \begin{subfigure}[t]{\columnwidth}
        \centering
        \includegraphics[width=\textwidth]{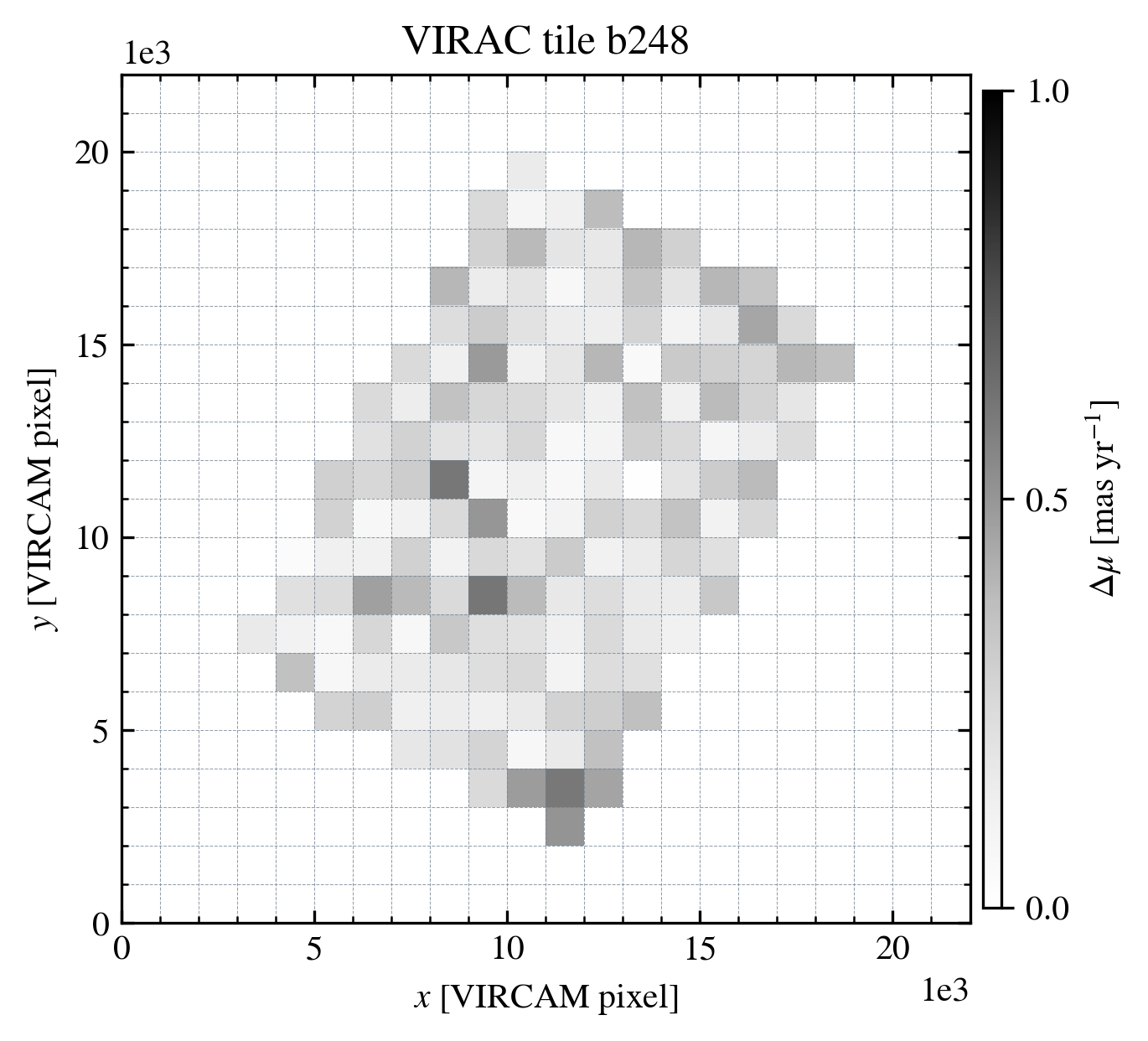}
    \end{subfigure}%
    ~
    \centering
    \begin{subfigure}[t]{\columnwidth}
        \centering
        \includegraphics[width=\textwidth]{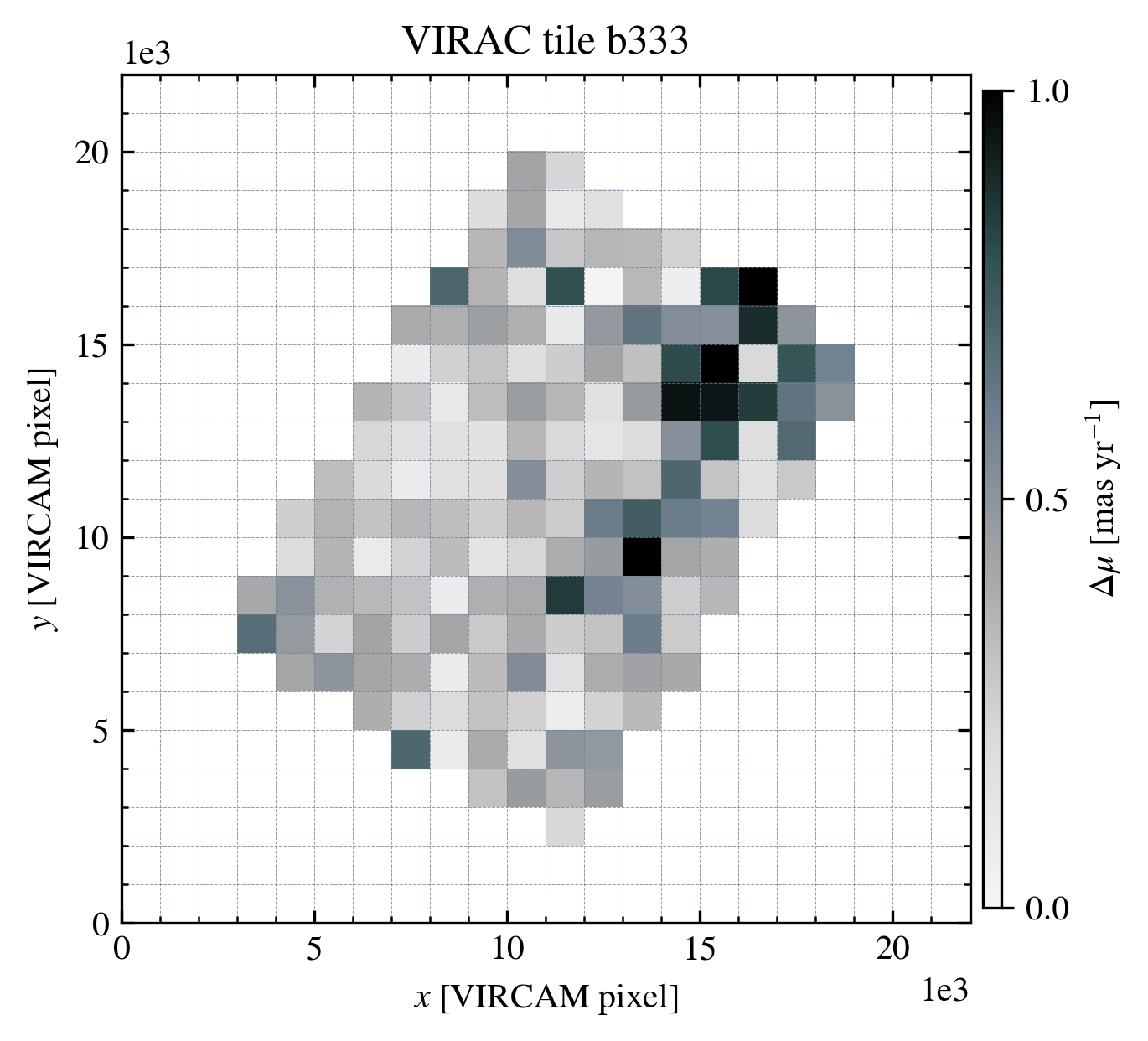}
    \end{subfigure}%
    \caption{Similar to Fig.\,\ref{fig:pms_comp_color}, but for {\it VIRAC} sources.}
    \label{fig:pms_comp_virac_color}
\end{figure*}

As a further cross-check, we compared our astrometry with the current public release of the \textit{VIRAC}
catalog \citep{virac}.
We cross-matched the \textit{Gaia} and \textit{VIRAC}
catalogs together, and performed the same comparisons that we described in the
previous section.
First, in Fig.\,\ref{fig:cmds_virac}, we show the color-magnitude
diagrams of the two tiles built using all the sources in the {\it VIRAC}
catalog. A comparison between their color-magnitude diagrams
and the ones in Fig.\,\ref{fig:cmds} shows
that our photometry is less affected by saturation effects.
Saturation affects sources with $K_{\rm S}\lesssim12$, and its effect
in the photometry is more evident in the color-magnitude diagrams of tile \texttt{b248},
where the \textit{VIRAC} photometry of the Bulge red giant branch exhibits a discontinuity.
The color-magnitude diagrams of tile \texttt{b333} also shows that our catalog
is deeper than \textit{VIRAC}.

Figure\,\ref{fig:pms_comp_virac} is similar to Fig.\,\ref{fig:pms_comp}, 
but for {\it VIRAC} sources. 
Dark grey points represent the sources with proper-motion error smaller than
2\,mas\,yr$^{-1}$ and with $12.5<K_{\rm S}<14$, and with the flag
\texttt{reliable}\,$=$\,1 in the \textit{VIRAC} catalog,
which exclude bad detections and poorly measured stars.
We notice that, apart from a global offset with respect to {\it Gaia},
the dispersions of the residuals are just slightly larger than ours. 
Since the formal proper motion errors given 
by the {\it VIRAC} catalog are marginally smaller than ours, 
we suspect that also {\it VIRAC} proper-motion errors might be underestimated.

Finally, in Fig.\,\ref{fig:pms_comp_virac_color},
we show the distribution of the residuals across the field of view,
for both tile \texttt{b248} (left) and \texttt{b333} (right).
Both tiles present significantly larger scatter 
compared to our work (cfr. Fig.\,\ref{fig:pms_comp_color}),
with systematic trends that depend on position.
Notice that {\it VIRAC}, conversely to our work, does not depend
on {\it Gaia},
and as such their systematic errors cannot be attributed
to the density of {\it Gaia} sources.
As pointed out in \cite{virac}, {\it VIRAC} proper motions are relative to
the average motion of sources within a few arcminutes, and in regions
with large spatial variations in the extinction, the bulk motion of
the reference stars can be different.\\

In summary, this work represents a step forward with respect to {\it VIRAC},
and the most notable improvements are:\ 
\textit{(i)} a larger number of sources with proper motions, mainly thanks to second-pass photometry; 
\textit{(ii)} significantly better astrometric precision (about 20-30\%) thanks to improved PSFs,
GD and image-motion correction, and \textit{(iii)} improved accuracy thanks to the registration
to the absolute reference system of \textit{Gaia}\,DR3, 
which was simply not available at the time of the {\it VIRAC} release.

As a final remark, our work uses independent tools from those of {\it VIRAC},
and can be leveraged for potential validations/benchmarks
for the upcoming {\it VIRAC}\, version 2 (Smith\,et\,al., in preparation).

\section{Parallax fit}
\label{sec:px}
\begin{figure*}
    \centering
    \begin{subfigure}[t]{0.38\textwidth}
        \centering
        \includegraphics[width=\textwidth]{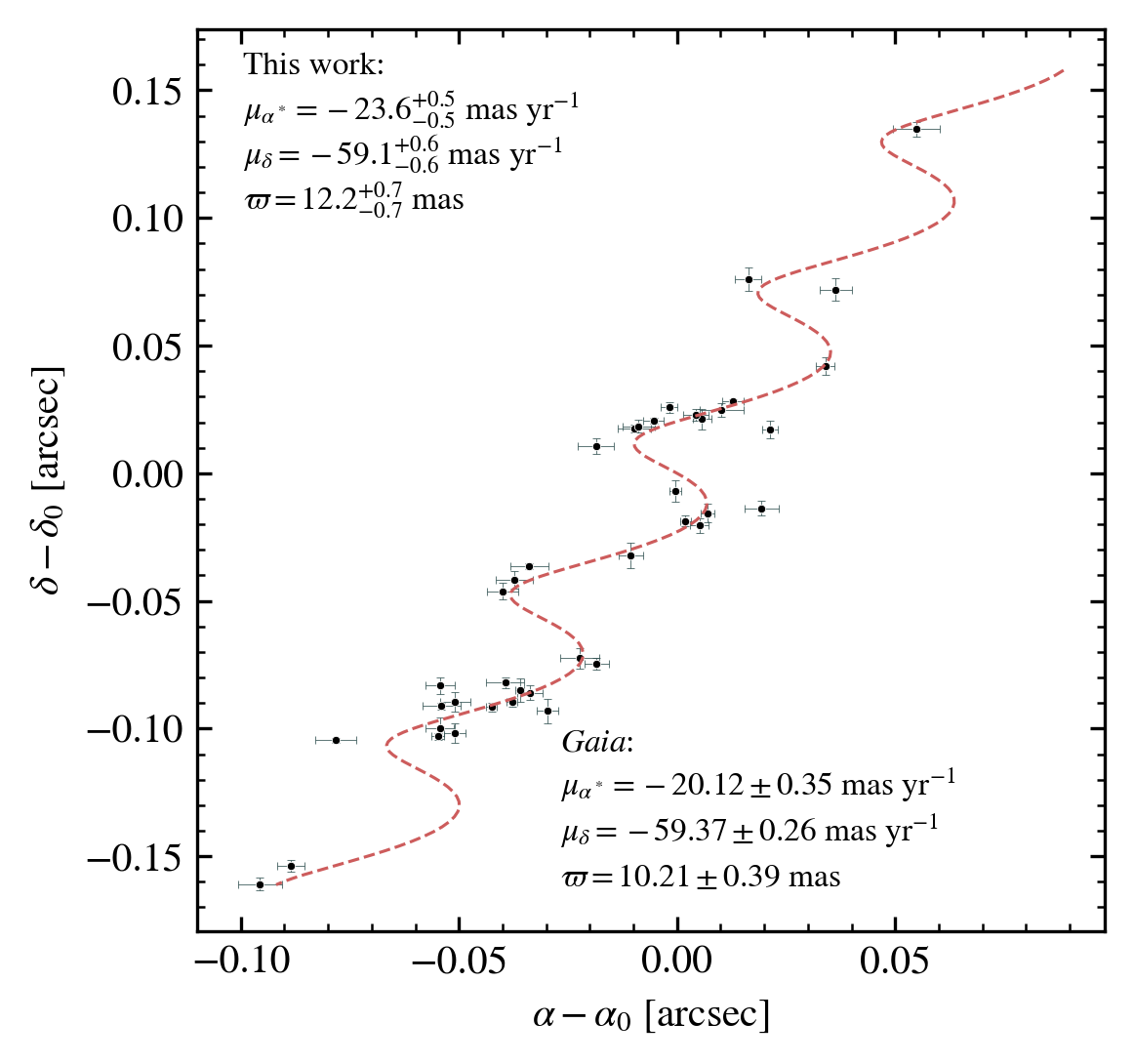}
        \caption{\textit{Gaia} DR3 source id \texttt{4042572202828764928}}
        \label{s1}
    \end{subfigure}%
    ~ 
    \centering
    \begin{subfigure}[t]{0.38\textwidth}
        \centering
        \includegraphics[width=\textwidth]{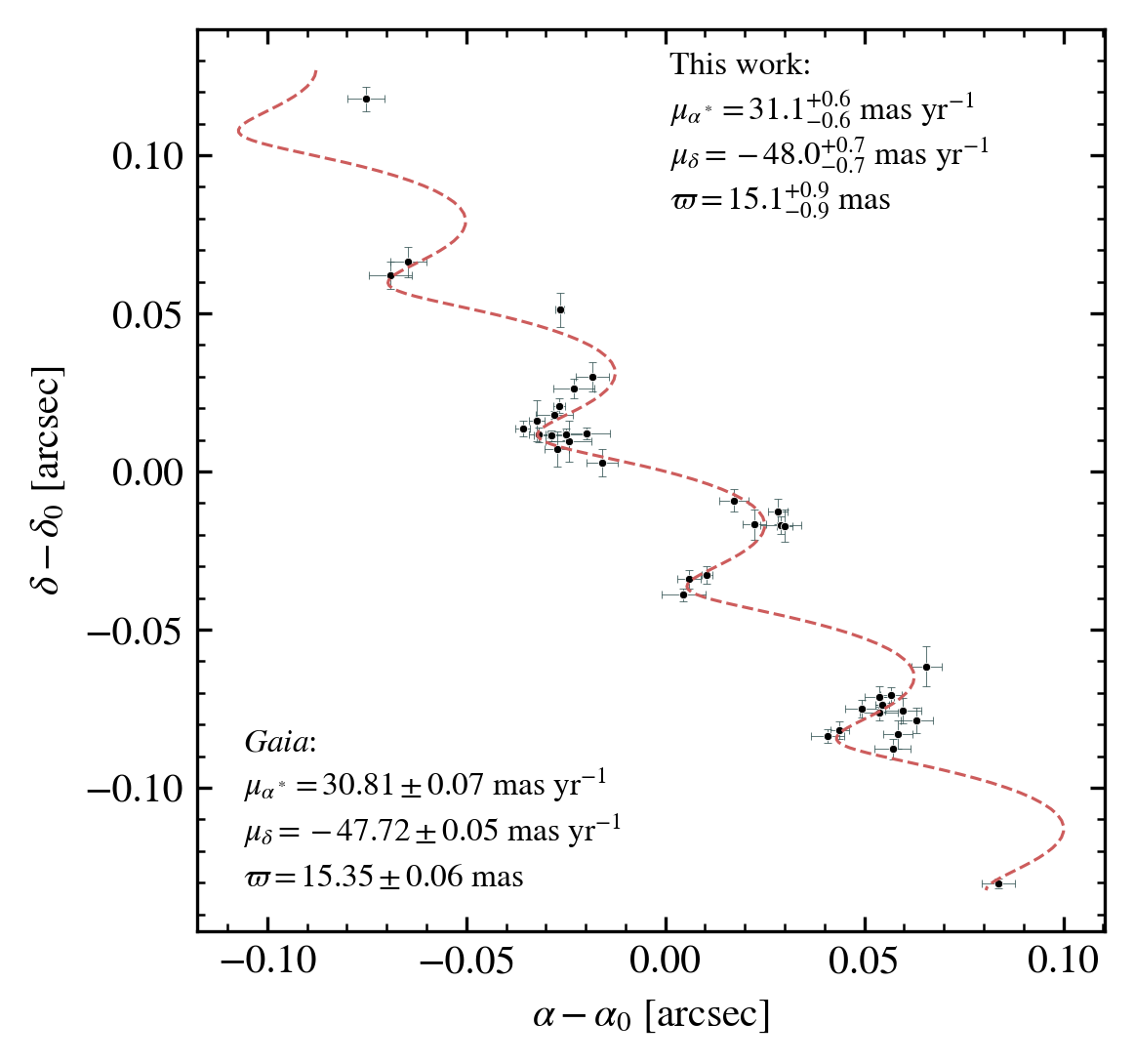}
        \caption{\textit{Gaia} DR3 source id \texttt{4039536902304599680}}
        \label{s2}
    \end{subfigure}%
    \\
    ~
    \centering
    \begin{subfigure}[t]{0.38\textwidth}
        \centering
        \includegraphics[width=\textwidth]{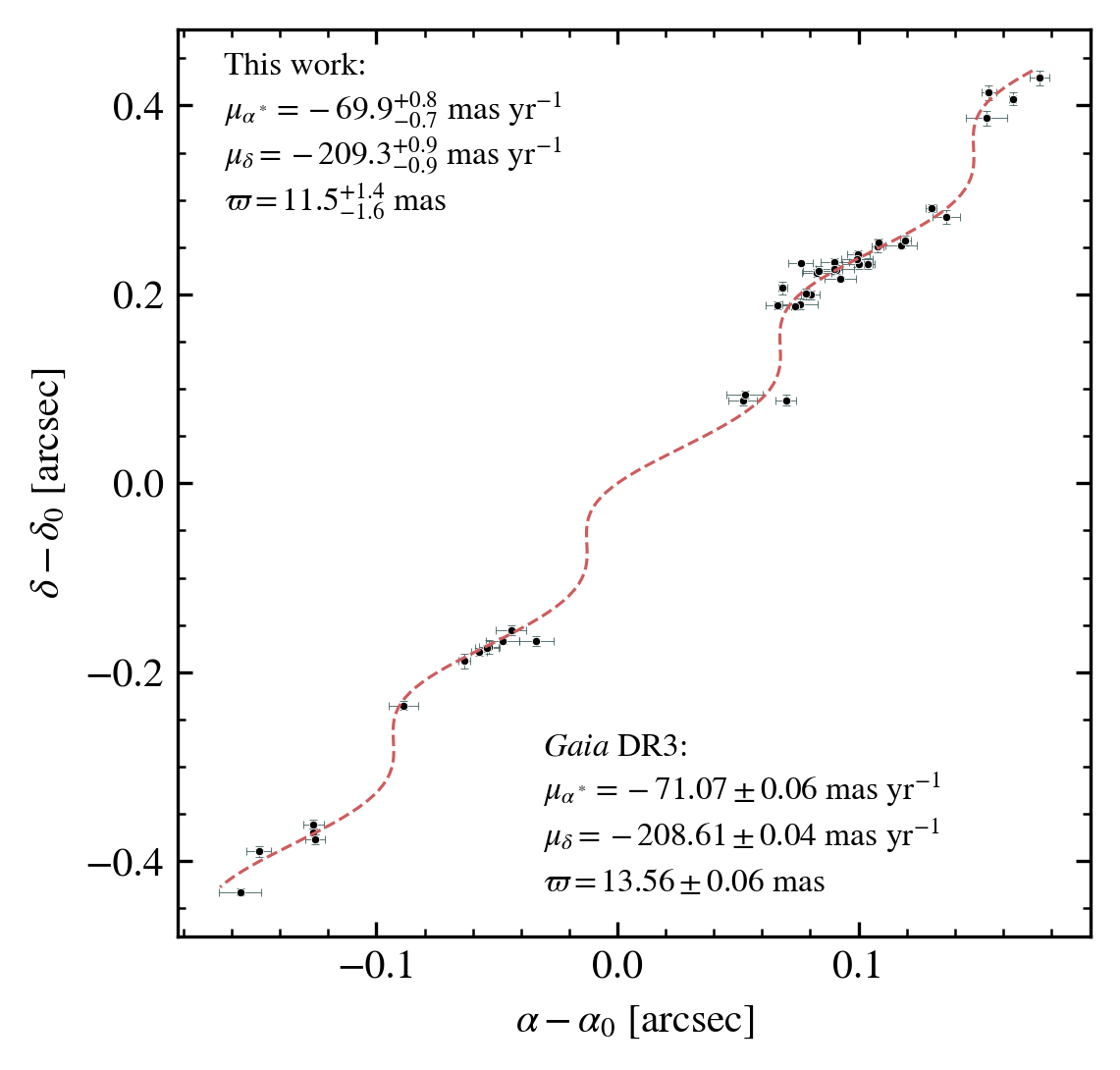}
        \caption{\textit{Gaia} DR3 source id \texttt{4057095072096565632}}
        \label{s3}
    \end{subfigure}%
    ~ 
    \centering
    \begin{subfigure}[t]{0.38\textwidth}
        \centering
        \includegraphics[width=\textwidth]{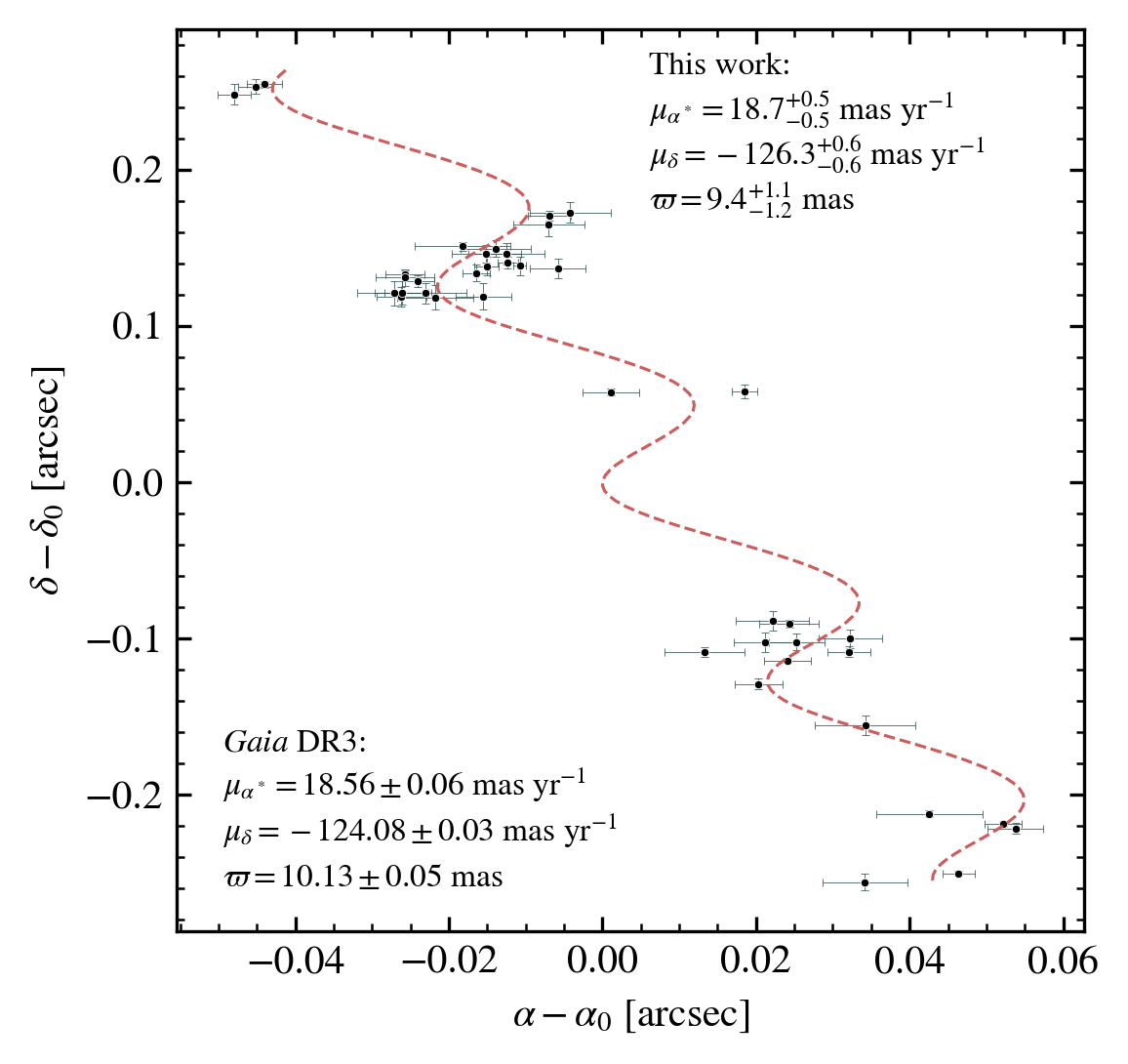}
        \caption{\textit{Gaia} DR3 source id \texttt{4056967490090373760}}
        \label{s4}
    \end{subfigure}%
    \caption{Example of parallax fit for four sources in common with \textit{Gaia}: panels \ref{s1} and \ref{s2} are for sources in tile \texttt{b248},
    while sources in panels \ref{s3} and \ref{s4} are in tile \texttt{b333}.
    Coordinates are relative to $(\alpha_0,\delta_0)$ chosen as the mean
    position between the first and last epochs.}
    \label{fig:px_fit}
\end{figure*}

The relatively high cadence of observations within the VVV dataset also enables the measurement of parallaxes,
at least for sources close enough to the Sun,
so that the effect of the parallax on their apparent motion can be effectively
disentangled from that of random positional errors.
We tested our parallax fitting procedure on a sample of sources in common with \textit{Gaia};
in particular, we selected well-measured sources with positions determined in at
least 20 epochs, a time baseline of at least 3 years, a proper motion error 
smaller than 2\,mas\,yr$^{-1}$ and with \textit{Gaia} parallax larger than 5\,mas (distance $<200$ pc).
The parallax fit was performed using the \texttt{Python NOVAS} libraries \citep{novas},
adopting the same procedure as in \cite{2017MNRAS.470.1140B}.
Briefly, we calculated the \texttt{NOVAS} predicted
source positions at each epoch for a given $(\alpha_{2000},\delta_{2000})$,
proper motion and parallax, where $(\alpha_{2000},\delta_{2000})$ are the positions at epoch 2000.0.
We then computed the differences between these positions
and those measured by us, and we look for the astrometric
solution that minimizes these differences, using the same MCMC approach that we employed
for the proper motion fit. 

In Fig.\,\ref{fig:px_fit}, we show an example of the parallax fit,
for four selected sources (two per tile).
We report in the plot the fitted proper motion and parallax values, together with
the values given in the {\it Gaia} DR3 catalog. Sources in Figs. \ref{s1} and \ref{s3} are 
in agreement within $2\sigma$, while those in Figs. \ref{s2} and \ref{s4} are compatible within
$1\sigma$ with the {\it Gaia} parallaxes.

\section{Data reduction strategy and access}
\label{sec:dra}
We plan to process the entire VVV dataset starting from the innermost Bulge fields.
However, we will also accept requests from the astronomical community to prioritize
particular VVV/VVVx tiles.
The catalogs of the first two tiles presented in this work are made available at the url
\texttt{\url{https://web.oapd.inaf.it/bedin/files/PAPERs_eMATERIALs/VVV-VVVx/}}. 
This repository will be constantly updated
with new products once they are ready.
For reasonable requests, we can also provide artificial star tests and astro-photometric time series.

\section{Conclusion}
\label{sec:concl}
In this paper, we exploited the VVV data to extend the {\it Gaia} astrometry into the Galactic plane, focusing on two pilot
fields, one in the Galactic center and the other in the South-East Bulge. 
We were able to significantly improve astrometric precision and completeness with respect to previous efforts.
These improvements are achieved through a combination of state-of-the-art techniques: (i) the use of spatially variable ePSFs for precise position and flux measurements; (ii) local transformations, which allow us to mitigate systematic errors, most notably residual geometric distortion and atmospheric effects; (iii) a combination of first- and second-pass photometry to improve the detection of faint stars in crowded fields, which allowed to
detect significantly more sources than previous efforts in the literature.
Our astrometry is anchored to the {\it Gaia} DR3 reference frame,
and represents an extension of the {\it Gaia}
accuracy into the Galactic plane.
In future releases, we plan to also include the VVVx data
to increase the number of pawprints and extend the temporal baseline available for each tile.
We will also combine the information from adjacent tiles in order to measure accurate proper motions also for sources near the borders of the tiles.

\begin{acknowledgements}
MG and AB acknowledge support by STScI DRF D0001.82523. MG and LRB acknowledge support by MIUR under PRIN program \#2017Z2HSMF and by PRIN-INAF 2019 under program \#10-Bedin.
DM gratefully acknowledges support from the ANID BASAL projects ACE210002 and FB210003,
from Fondecyt Project No. 1220724, and from CNPq Brasil Project 350104/2022-0.
The authors thank Maren Hempel for providing the VVV tiles pointing data.
\end{acknowledgements}


%
%
\bibliographystyle{aa} 
\bibliography{bibliography} 

\begin{thebibliography}{45}
\expandafter\ifx\csname natexlab\endcsname\relax\def\natexlab#1{#1}\fi

\bibitem[{{Anderson} {et~al.}(2006){Anderson}, {Bedin}, {Piotto}, {Yadav}, \&
  {Bellini}}]{2006A&A...454.1029A}
{Anderson}, J., {Bedin}, L.~R., {Piotto}, G., {Yadav}, R.~S., \& {Bellini}, A.
  2006, \aap, 454, 1029

\bibitem[{{Anderson} {et~al.}(2008{\natexlab{a}}){Anderson}, {King}, {Richer},
  {Fahlman}, {Hansen}, {Hurley}, {Kalirai}, {Rich}, \&
  {Stetson}}]{2008AJ....135.2114A}
{Anderson}, J., {King}, I.~R., {Richer}, H.~B., {et~al.} 2008{\natexlab{a}},
  \aj, 135, 2114

\bibitem[{{Anderson} {et~al.}(2008{\natexlab{b}}){Anderson}, {Sarajedini},
  {Bedin}, {King}, {Piotto}, {Reid}, {Siegel}, {Majewski}, {Paust}, {Aparicio},
  {Milone}, {Chaboyer}, \& {Rosenberg}}]{2008AJ....135.2055A}
{Anderson}, J., {Sarajedini}, A., {Bedin}, L.~R., {et~al.} 2008{\natexlab{b}},
  \aj, 135, 2055

\bibitem[{{Anderson} \& {van der Marel}(2010)}]{2010ApJ...710.1032A}
{Anderson}, J. \& {van der Marel}, R.~P. 2010, \apj, 710, 1032

\bibitem[{{Barbuy} {et~al.}(2018){Barbuy}, {Chiappini}, \&
  {Gerhard}}]{2018ARA&A..56..223B}
{Barbuy}, B., {Chiappini}, C., \& {Gerhard}, O. 2018, \araa, 56, 223

\bibitem[{{Barron} {et~al.}(2011){Barron}, {Kaplan}, {Bangert}, {Bartlett},
  {Puatua}, {Harris}, \& {Barrett}}]{novas}
{Barron}, E.~G., {Kaplan}, G.~H., {Bangert}, J., {et~al.} 2011, in American
  Astronomical Society Meeting Abstracts, Vol. 217, American Astronomical
  Society Meeting Abstracts \#217, 344.14

\bibitem[{{Bedin} \& {Fontanive}(2018)}]{2018MNRAS.481.5339B}
{Bedin}, L.~R. \& {Fontanive}, C. 2018, \mnras, 481, 5339

\bibitem[{{Bedin} \& {Fontanive}(2020)}]{2020MNRAS.494.2068B}
{Bedin}, L.~R. \& {Fontanive}, C. 2020, \mnras, 494, 2068

\bibitem[{{Bedin} {et~al.}(2017){Bedin}, {Pourbaix}, {Apai}, {Burgasser},
  {Buenzli}, {Boffin}, \& {Libralato}}]{2017MNRAS.470.1140B}
{Bedin}, L.~R., {Pourbaix}, D., {Apai}, D., {et~al.} 2017, \mnras, 470, 1140

\bibitem[{{Bedin} {et~al.}(2014){Bedin}, {Ruiz-Lapuente}, {Gonz{\'a}lez
  Hern{\'a}ndez}, {Canal}, {Filippenko}, \& {Mendez}}]{2014MNRAS.439..354B}
{Bedin}, L.~R., {Ruiz-Lapuente}, P., {Gonz{\'a}lez Hern{\'a}ndez}, J.~I.,
  {et~al.} 2014, \mnras, 439, 354

\bibitem[{{Bedin} {et~al.}(2009){Bedin}, {Salaris}, {Piotto}, {Anderson},
  {King}, \& {Cassisi}}]{2009ApJ...697..965B}
{Bedin}, L.~R., {Salaris}, M., {Piotto}, G., {et~al.} 2009, \apj, 697, 965

\bibitem[{{Bellini} {et~al.}(2017){Bellini}, {Anderson}, {Bedin}, {King}, {van
  der Marel}, {Piotto}, \& {Cool}}]{2017ApJ...842....6B}
{Bellini}, A., {Anderson}, J., {Bedin}, L.~R., {et~al.} 2017, \apj, 842, 6

\bibitem[{{Contreras Ramos} {et~al.}(2018){Contreras Ramos}, {Minniti},
  {Fern{\'a}ndez-Trincado}, {Alonso-Garc{\'\i}a}, {Catelan}, {Gran}, {Hajdu},
  {Hanke}, {Hempel}, {Moreno D{\'\i}az}, {P{\'e}rez-Villegas},
  {Rojas-Arriagada}, \& {Zoccali}}]{2018ApJ...863...78C}
{Contreras Ramos}, R., {Minniti}, D., {Fern{\'a}ndez-Trincado}, J.~G., {et~al.}
  2018, \apj, 863, 78

\bibitem[{{Fabricius} {et~al.}(2021){Fabricius}, {Luri}, {Arenou}, {Babusiaux},
  {Helmi}, {Muraveva}, {Reyl{\'e}}, {Spoto}, {Vallenari}, {Antoja}, {Balbinot},
  {Barache}, {Bauchet}, {Bragaglia}, {Busonero}, {Cantat-Gaudin}, {Carrasco},
  {Diakit{\'e}}, {Fabrizio}, {Figueras}, {Garcia-Gutierrez}, {Garofalo},
  {Jordi}, {Kervella}, {Khanna}, {Leclerc}, {Licata}, {Lambert}, {Marrese},
  {Masip}, {Ramos}, {Robichon}, {Robin}, {Romero-G{\'o}mez}, {Rubele}, \&
  {Weiler}}]{2021A&A...649A...5F}
{Fabricius}, C., {Luri}, X., {Arenou}, F., {et~al.} 2021, \aap, 649, A5

\bibitem[{{Foreman-Mackey} {et~al.}(2013){Foreman-Mackey}, {Hogg}, {Lang}, \&
  {Goodman}}]{2013PASP..125..306F}
{Foreman-Mackey}, D., {Hogg}, D.~W., {Lang}, D., \& {Goodman}, J. 2013, \pasp,
  125, 306

\bibitem[{{Fragkoudi} {et~al.}(2020){Fragkoudi}, {Di Matteo}, {Haywood},
  {Khoperskov}, {Gomez}, {Schultheis}, {Combes}, \&
  {Semelin}}]{2020IAUGA..30..282F}
{Fragkoudi}, F., {Di Matteo}, P., {Haywood}, M., {et~al.} 2020, in IAU General
  Assembly, 282--283

\bibitem[{{Gaia Collaboration} {et~al.}(2016){Gaia Collaboration}, {Prusti},
  {de Bruijne}, {Brown}, {Vallenari}, {Babusiaux}, {Bailer-Jones}, {Bastian},
  {Biermann}, {Evans}, {Eyer}, {Jansen}, {Jordi}, {Klioner}, {Lammers},
  {Lindegren}, {Luri}, {Mignard}, {Milligan}, {Panem}, {Poinsignon},
  {Pourbaix}, {Randich}, {Sarri}, {Sartoretti}, {Siddiqui}, {Soubiran},
  {Valette}, {van Leeuwen}, {Walton}, {Aerts}, {Arenou}, {Cropper}, {Drimmel},
  {H{\o}g}, {Katz}, {Lattanzi}, {O'Mullane}, {Grebel}, {Holland}, {Huc},
  {Passot}, {Bramante}, {Cacciari}, {Casta{\~n}eda}, {Chaoul}, {Cheek}, {De
  Angeli}, {Fabricius}, {Guerra}, {Hern{\'a}ndez}, {Jean-Antoine-Piccolo},
  {Masana}, {Messineo}, {Mowlavi}, {Nienartowicz}, {Ord{\'o}{\~n}ez-Blanco},
  {Panuzzo}, {Portell}, {Richards}, {Riello}, {Seabroke}, {Tanga},
  {Th{\'e}venin}, {Torra}, {Els}, {Gracia-Abril}, {Comoretto},
  {Garcia-Reinaldos}, {Lock}, {Mercier}, {Altmann}, {Andrae}, {Astraatmadja},
  {Bellas-Velidis}, {Benson}, {Berthier}, {Blomme}, {Busso}, {Carry},
  {Cellino}, {Clementini}, {Cowell}, {Creevey}, {Cuypers}, {Davidson}, {De
  Ridder}, {de Torres}, {Delchambre}, {Dell'Oro}, {Ducourant}, {Fr{\'e}mat},
  {Garc{\'\i}a-Torres}, {Gosset}, {Halbwachs}, {Hambly}, {Harrison}, {Hauser},
  {Hestroffer}, {Hodgkin}, {Huckle}, {Hutton}, {Jasniewicz}, {Jordan},
  {Kontizas}, {Korn}, {Lanzafame}, {Manteiga}, {Moitinho}, {Muinonen},
  {Osinde}, {Pancino}, {Pauwels}, {Petit}, {Recio-Blanco}, {Robin}, {Sarro},
  {Siopis}, {Smith}, {Smith}, {Sozzetti}, {Thuillot}, {van Reeven}, {Viala},
  {Abbas}, {Abreu Aramburu}, {Accart}, {Aguado}, {Allan}, {Allasia},
  {Altavilla}, {{\'A}lvarez}, {Alves}, {Anderson}, {Andrei}, {Anglada Varela},
  {Antiche}, {Antoja}, {Ant{\'o}n}, {Arcay}, {Atzei}, {Ayache}, {Bach},
  {Baker}, {Balaguer-N{\'u}{\~n}ez}, {Barache}, {Barata}, {Barbier}, {Barblan},
  {Baroni}, {Barrado y Navascu{\'e}s}, {Barros}, {Barstow}, {Becciani},
  {Bellazzini}, {Bellei}, {Bello Garc{\'\i}a}, {Belokurov}, {Bendjoya},
  {Berihuete}, {Bianchi}, {Bienaym{\'e}}, {Billebaud}, {Blagorodnova},
  {Blanco-Cuaresma}, {Boch}, {Bombrun}, {Borrachero}, {Bouquillon}, {Bourda},
  {Bouy}, {Bragaglia}, {Breddels}, {Brouillet}, {Br{\"u}semeister},
  {Bucciarelli}, {Budnik}, {Burgess}, {Burgon}, {Burlacu}, {Busonero}, {Buzzi},
  {Caffau}, {Cambras}, {Campbell}, {Cancelliere}, {Cantat-Gaudin}, {Carlucci},
  {Carrasco}, {Castellani}, {Charlot}, {Charnas}, {Charvet}, {Chassat},
  {Chiavassa}, {Clotet}, {Cocozza}, {Collins}, {Collins}, {Costigan}, {Crifo},
  {Cross}, {Crosta}, {Crowley}, {Dafonte}, {Damerdji}, {Dapergolas}, {David},
  {David}, {De Cat}, {de Felice}, {de Laverny}, {De Luise}, {De March}, {de
  Martino}, {de Souza}, {Debosscher}, {del Pozo}, {Delbo}, {Delgado},
  {Delgado}, {di Marco}, {Di Matteo}, {Diakite}, {Distefano}, {Dolding}, {Dos
  Anjos}, {Drazinos}, {Dur{\'a}n}, {Dzigan}, {Ecale}, {Edvardsson}, {Enke},
  {Erdmann}, {Escolar}, {Espina}, {Evans}, {Eynard Bontemps}, {Fabre},
  {Fabrizio}, {Faigler}, {Falc{\~a}o}, {Farr{\`a}s Casas}, {Faye}, {Federici},
  {Fedorets}, {Fern{\'a}ndez-Hern{\'a}ndez}, {Fernique}, {Fienga}, {Figueras},
  {Filippi}, {Findeisen}, {Fonti}, {Fouesneau}, {Fraile}, {Fraser}, {Fuchs},
  {Furnell}, {Gai}, {Galleti}, {Galluccio}, {Garabato}, {Garc{\'\i}a-Sedano},
  {Gar{\'e}}, {Garofalo}, {Garralda}, {Gavras}, {Gerssen}, {Geyer}, {Gilmore},
  {Girona}, {Giuffrida}, {Gomes}, {Gonz{\'a}lez-Marcos},
  {Gonz{\'a}lez-N{\'u}{\~n}ez}, {Gonz{\'a}lez-Vidal}, {Granvik}, {Guerrier},
  {Guillout}, {Guiraud}, {G{\'u}rpide}, {Guti{\'e}rrez-S{\'a}nchez}, {Guy},
  {Haigron}, {Hatzidimitriou}, {Haywood}, {Heiter}, {Helmi}, {Hobbs},
  {Hofmann}, {Holl}, {Holland}, {Hunt}, {Hypki}, {Icardi}, {Irwin}, {Jevardat
  de Fombelle}, {Jofr{\'e}}, {Jonker}, {Jorissen}, {Julbe}, {Karampelas},
  {Kochoska}, {Kohley}, {Kolenberg}, {Kontizas}, {Koposov}, {Kordopatis},
  {Koubsky}, {Kowalczyk}, {Krone-Martins}, {Kudryashova}, {Kull}, {Bachchan},
  {Lacoste-Seris}, {Lanza}, {Lavigne}, {Le Poncin-Lafitte}, {Lebreton},
  {Lebzelter}, {Leccia}, {Leclerc}, {Lecoeur-Taibi}, {Lemaitre}, {Lenhardt},
  {Leroux}, {Liao}, {Licata}, {Lindstr{\o}m}, {Lister}, {Livanou}, {Lobel},
  {L{\"o}ffler}, {L{\'o}pez}, {Lopez-Lozano}, {Lorenz}, {Loureiro},
  {MacDonald}, {Magalh{\~a}es Fernandes}, {Managau}, {Mann}, {Mantelet},
  {Marchal}, {Marchant}, {Marconi}, {Marie}, {Marinoni}, {Marrese},
  {Marschalk{\'o}}, {Marshall}, {Mart{\'\i}n-Fleitas}, {Martino}, {Mary},
  {Matijevi{\v{c}}}, {Mazeh}, {McMillan}, {Messina}, {Mestre}, {Michalik},
  {Millar}, {Miranda}, {Molina}, {Molinaro}, {Molinaro}, {Moln{\'a}r},
  {Moniez}, {Montegriffo}, {Monteiro}, {Mor}, {Mora}, {Morbidelli}, {Morel},
  {Morgenthaler}, {Morley}, {Morris}, {Mulone}, {Muraveva}, {Musella},
  {Narbonne}, {Nelemans}, {Nicastro}, {Noval}, {Ord{\'e}novic},
  {Ordieres-Mer{\'e}}, {Osborne}, {Pagani}, {Pagano}, {Pailler}, {Palacin},
  {Palaversa}, {Parsons}, {Paulsen}, {Pecoraro}, {Pedrosa}, {Pentik{\"a}inen},
  {Pereira}, {Pichon}, {Piersimoni}, {Pineau}, {Plachy}, {Plum}, {Poujoulet},
  {Pr{\v{s}}a}, {Pulone}, {Ragaini}, {Rago}, {Rambaux}, {Ramos-Lerate},
  {Ranalli}, {Rauw}, {Read}, {Regibo}, {Renk}, {Reyl{\'e}}, {Ribeiro},
  {Rimoldini}, {Ripepi}, {Riva}, {Rixon}, {Roelens}, {Romero-G{\'o}mez},
  {Rowell}, {Royer}, {Rudolph}, {Ruiz-Dern}, {Sadowski}, {Sagrist{\`a}
  Sell{\'e}s}, {Sahlmann}, {Salgado}, {Salguero}, {Sarasso}, {Savietto},
  {Schnorhk}, {Schultheis}, {Sciacca}, {Segol}, {Segovia}, {Segransan},
  {Serpell}, {Shih}, {Smareglia}, {Smart}, {Smith}, {Solano}, {Solitro},
  {Sordo}, {Soria Nieto}, {Souchay}, {Spagna}, {Spoto}, {Stampa}, {Steele},
  {Steidelm{\"u}ller}, {Stephenson}, {Stoev}, {Suess}, {S{\"u}veges}, {Surdej},
  {Szabados}, {Szegedi-Elek}, {Tapiador}, {Taris}, {Tauran}, {Taylor},
  {Teixeira}, {Terrett}, {Tingley}, {Trager}, {Turon}, {Ulla}, {Utrilla},
  {Valentini}, {van Elteren}, {Van Hemelryck}, {van Leeuwen}, {Varadi},
  {Vecchiato}, {Veljanoski}, {Via}, {Vicente}, {Vogt}, {Voss}, {Votruba},
  {Voutsinas}, {Walmsley}, {Weiler}, {Weingrill}, {Werner}, {Wevers},
  {Whitehead}, {Wyrzykowski}, {Yoldas}, {{\v{Z}}erjal}, {Zucker}, {Zurbach},
  {Zwitter}, {Alecu}, {Allen}, {Allende Prieto}, {Amorim},
  {Anglada-Escud{\'e}}, {Arsenijevic}, {Azaz}, {Balm}, {Beck}, {Bernstein},
  {Bigot}, {Bijaoui}, {Blasco}, {Bonfigli}, {Bono}, {Boudreault}, {Bressan},
  {Brown}, {Brunet}, {Bunclark}, {Buonanno}, {Butkevich}, {Carret}, {Carrion},
  {Chemin}, {Ch{\'e}reau}, {Corcione}, {Darmigny}, {de Boer}, {de Teodoro}, {de
  Zeeuw}, {Delle Luche}, {Domingues}, {Dubath}, {Fodor}, {Fr{\'e}zouls},
  {Fries}, {Fustes}, {Fyfe}, {Gallardo}, {Gallegos}, {Gardiol}, {Gebran},
  {Gomboc}, {G{\'o}mez}, {Grux}, {Gueguen}, {Heyrovsky}, {Hoar}, {Iannicola},
  {Isasi Parache}, {Janotto}, {Joliet}, {Jonckheere}, {Keil}, {Kim},
  {Klagyivik}, {Klar}, {Knude}, {Kochukhov}, {Kolka}, {Kos}, {Kutka}, {Lainey},
  {LeBouquin}, {Liu}, {Loreggia}, {Makarov}, {Marseille}, {Martayan},
  {Martinez-Rubi}, {Massart}, {Meynadier}, {Mignot}, {Munari}, {Nguyen},
  {Nordlander}, {Ocvirk}, {O'Flaherty}, {Olias Sanz}, {Ortiz}, {Osorio},
  {Oszkiewicz}, {Ouzounis}, {Palmer}, {Park}, {Pasquato}, {Peltzer}, {Peralta},
  {P{\'e}turaud}, {Pieniluoma}, {Pigozzi}, {Poels}, {Prat}, {Prod'homme},
  {Raison}, {Rebordao}, {Risquez}, {Rocca-Volmerange}, {Rosen}, {Ruiz-Fuertes},
  {Russo}, {Sembay}, {Serraller Vizcaino}, {Short}, {Siebert}, {Silva},
  {Sinachopoulos}, {Slezak}, {Soffel}, {Sosnowska}, {Strai{\v{z}}ys}, {ter
  Linden}, {Terrell}, {Theil}, {Tiede}, {Troisi}, {Tsalmantza}, {Tur},
  {Vaccari}, {Vachier}, {Valles}, {Van Hamme}, {Veltz}, {Virtanen}, {Wallut},
  {Wichmann}, {Wilkinson}, {Ziaeepour}, \& {Zschocke}}]{gaiamission}
{Gaia Collaboration}, {Prusti}, T., {de Bruijne}, J.~H.~J., {et~al.} 2016,
  \aap, 595, A1

\bibitem[{{Gaia Collaboration} {et~al.}(2023){Gaia Collaboration}, {Vallenari},
  {Brown}, {Prusti}, {de Bruijne}, {Arenou}, {Babusiaux}, {Biermann},
  {Creevey}, {Ducourant}, {Evans}, {Eyer}, {Guerra}, {Hutton}, {Jordi},
  {Klioner}, {Lammers}, {Lindegren}, {Luri}, {Mignard}, {Panem}, {Pourbaix},
  {Randich}, {Sartoretti}, {Soubiran}, {Tanga}, {Walton}, {Bailer-Jones},
  {Bastian}, {Drimmel}, {Jansen}, {Katz}, {Lattanzi}, {van Leeuwen}, {Bakker},
  {Cacciari}, {Casta{\~n}eda}, {De Angeli}, {Fabricius}, {Fouesneau},
  {Fr{\'e}mat}, {Galluccio}, {Guerrier}, {Heiter}, {Masana}, {Messineo},
  {Mowlavi}, {Nicolas}, {Nienartowicz}, {Pailler}, {Panuzzo}, {Riclet}, {Roux},
  {Seabroke}, {Sordo}, {Th{\'e}venin}, {Gracia-Abril}, {Portell}, {Teyssier},
  {Altmann}, {Andrae}, {Audard}, {Bellas-Velidis}, {Benson}, {Berthier},
  {Blomme}, {Burgess}, {Busonero}, {Busso}, {C{\'a}novas}, {Carry}, {Cellino},
  {Cheek}, {Clementini}, {Damerdji}, {Davidson}, {de Teodoro}, {Nu{\~n}ez
  Campos}, {Delchambre}, {Dell'Oro}, {Esquej}, {Fern{\'a}ndez-Hern{\'a}ndez},
  {Fraile}, {Garabato}, {Garc{\'\i}a-Lario}, {Gosset}, {Haigron}, {Halbwachs},
  {Hambly}, {Harrison}, {Hern{\'a}ndez}, {Hestroffer}, {Hodgkin}, {Holl},
  {Jan{\ss}en}, {Jevardat de Fombelle}, {Jordan}, {Krone-Martins}, {Lanzafame},
  {L{\"o}ffler}, {Marchal}, {Marrese}, {Moitinho}, {Muinonen}, {Osborne},
  {Pancino}, {Pauwels}, {Recio-Blanco}, {Reyl{\'e}}, {Riello}, {Rimoldini},
  {Roegiers}, {Rybizki}, {Sarro}, {Siopis}, {Smith}, {Sozzetti}, {Utrilla},
  {van Leeuwen}, {Abbas}, {{\'A}brah{\'a}m}, {Abreu Aramburu}, {Aerts},
  {Aguado}, {Ajaj}, {Aldea-Montero}, {Altavilla}, {{\'A}lvarez}, {Alves},
  {Anders}, {Anderson}, {Anglada Varela}, {Antoja}, {Baines}, {Baker},
  {Balaguer-N{\'u}{\~n}ez}, {Balbinot}, {Balog}, {Barache}, {Barbato},
  {Barros}, {Barstow}, {Bartolom{\'e}}, {Bassilana}, {Bauchet}, {Becciani},
  {Bellazzini}, {Berihuete}, {Bernet}, {Bertone}, {Bianchi}, {Binnenfeld},
  {Blanco-Cuaresma}, {Blazere}, {Boch}, {Bombrun}, {Bossini}, {Bouquillon},
  {Bragaglia}, {Bramante}, {Breedt}, {Bressan}, {Brouillet}, {Brugaletta},
  {Bucciarelli}, {Burlacu}, {Butkevich}, {Buzzi}, {Caffau}, {Cancelliere},
  {Cantat-Gaudin}, {Carballo}, {Carlucci}, {Carnerero}, {Carrasco},
  {Casamiquela}, {Castellani}, {Castro-Ginard}, {Chaoul}, {Charlot}, {Chemin},
  {Chiaramida}, {Chiavassa}, {Chornay}, {Comoretto}, {Contursi}, {Cooper},
  {Cornez}, {Cowell}, {Crifo}, {Cropper}, {Crosta}, {Crowley}, {Dafonte},
  {Dapergolas}, {David}, {David}, {de Laverny}, {De Luise}, {De March}, {De
  Ridder}, {de Souza}, {de Torres}, {del Peloso}, {del Pozo}, {Delbo},
  {Delgado}, {Delisle}, {Demouchy}, {Dharmawardena}, {Di Matteo}, {Diakite},
  {Diener}, {Distefano}, {Dolding}, {Edvardsson}, {Enke}, {Fabre}, {Fabrizio},
  {Faigler}, {Fedorets}, {Fernique}, {Fienga}, {Figueras}, {Fournier},
  {Fouron}, {Fragkoudi}, {Gai}, {Garcia-Gutierrez}, {Garcia-Reinaldos},
  {Garc{\'\i}a-Torres}, {Garofalo}, {Gavel}, {Gavras}, {Gerlach}, {Geyer},
  {Giacobbe}, {Gilmore}, {Girona}, {Giuffrida}, {Gomel}, {Gomez},
  {Gonz{\'a}lez-N{\'u}{\~n}ez}, {Gonz{\'a}lez-Santamar{\'\i}a},
  {Gonz{\'a}lez-Vidal}, {Granvik}, {Guillout}, {Guiraud},
  {Guti{\'e}rrez-S{\'a}nchez}, {Guy}, {Hatzidimitriou}, {Hauser}, {Haywood},
  {Helmer}, {Helmi}, {Sarmiento}, {Hidalgo}, {Hilger}, {H{\l}adczuk}, {Hobbs},
  {Holland}, {Huckle}, {Jardine}, {Jasniewicz}, {Jean-Antoine Piccolo},
  {Jim{\'e}nez-Arranz}, {Jorissen}, {Juaristi Campillo}, {Julbe}, {Karbevska},
  {Kervella}, {Khanna}, {Kontizas}, {Kordopatis}, {Korn}, {K{\'o}sp{\'a}l},
  {Kostrzewa-Rutkowska}, {Kruszy{\'n}ska}, {Kun}, {Laizeau}, {Lambert},
  {Lanza}, {Lasne}, {Le Campion}, {Lebreton}, {Lebzelter}, {Leccia}, {Leclerc},
  {Lecoeur-Taibi}, {Liao}, {Licata}, {Lindstr{\o}m}, {Lister}, {Livanou},
  {Lobel}, {Lorca}, {Loup}, {Madrero Pardo}, {Magdaleno Romeo}, {Managau},
  {Mann}, {Manteiga}, {Marchant}, {Marconi}, {Marcos}, {Marcos Santos},
  {Mar{\'\i}n Pina}, {Marinoni}, {Marocco}, {Marshall}, {Martin Polo},
  {Mart{\'\i}n-Fleitas}, {Marton}, {Mary}, {Masip}, {Massari},
  {Mastrobuono-Battisti}, {Mazeh}, {McMillan}, {Messina}, {Michalik}, {Millar},
  {Mints}, {Molina}, {Molinaro}, {Moln{\'a}r}, {Monari}, {Mongui{\'o}},
  {Montegriffo}, {Montero}, {Mor}, {Mora}, {Morbidelli}, {Morel}, {Morris},
  {Muraveva}, {Murphy}, {Musella}, {Nagy}, {Noval}, {Oca{\~n}a}, {Ogden},
  {Ordenovic}, {Osinde}, {Pagani}, {Pagano}, {Palaversa}, {Palicio},
  {Pallas-Quintela}, {Panahi}, {Payne-Wardenaar}, {Pe{\~n}alosa Esteller},
  {Penttil{\"a}}, {Pichon}, {Piersimoni}, {Pineau}, {Plachy}, {Plum}, {Poggio},
  {Pr{\v{s}}a}, {Pulone}, {Racero}, {Ragaini}, {Rainer}, {Raiteri}, {Rambaux},
  {Ramos}, {Ramos-Lerate}, {Re Fiorentin}, {Regibo}, {Richards}, {Rios Diaz},
  {Ripepi}, {Riva}, {Rix}, {Rixon}, {Robichon}, {Robin}, {Robin}, {Roelens},
  {Rogues}, {Rohrbasser}, {Romero-G{\'o}mez}, {Rowell}, {Royer}, {Ruz Mieres},
  {Rybicki}, {Sadowski}, {S{\'a}ez N{\'u}{\~n}ez}, {Sagrist{\`a} Sell{\'e}s},
  {Sahlmann}, {Salguero}, {Samaras}, {Sanchez Gimenez}, {Sanna},
  {Santove{\~n}a}, {Sarasso}, {Schultheis}, {Sciacca}, {Segol}, {Segovia},
  {S{\'e}gransan}, {Semeux}, {Shahaf}, {Siddiqui}, {Siebert}, {Siltala},
  {Silvelo}, {Slezak}, {Slezak}, {Smart}, {Snaith}, {Solano}, {Solitro},
  {Souami}, {Souchay}, {Spagna}, {Spina}, {Spoto}, {Steele},
  {Steidelm{\"u}ller}, {Stephenson}, {S{\"u}veges}, {Surdej}, {Szabados},
  {Szegedi-Elek}, {Taris}, {Taylor}, {Teixeira}, {Tolomei}, {Tonello}, {Torra},
  {Torra}, {Torralba Elipe}, {Trabucchi}, {Tsounis}, {Turon}, {Ulla}, {Unger},
  {Vaillant}, {van Dillen}, {van Reeven}, {Vanel}, {Vecchiato}, {Viala},
  {Vicente}, {Voutsinas}, {Weiler}, {Wevers}, {Wyrzykowski}, {Yoldas}, {Yvard},
  {Zhao}, {Zorec}, {Zucker}, \& {Zwitter}}]{gaiadr3}
{Gaia Collaboration}, {Vallenari}, A., {Brown}, A.~G.~A., {et~al.} 2023, \aap,
  674, A1

\bibitem[{{Garro} {et~al.}(2022{\natexlab{a}}){Garro}, {Minniti}, {Alessi},
  {Patchick}, {Kronberger}, {Alonso-Garc{\'\i}a}, {Fern{\'a}ndez-Trincado},
  {G{\'o}mez}, {Hempel}, {Pullen}, {Saito}, {Ripepi}, \& {Zelada
  Bacigalupo}}]{2022A&A...659A.155G}
{Garro}, E.~R., {Minniti}, D., {Alessi}, B., {et~al.} 2022{\natexlab{a}}, \aap,
  659, A155

\bibitem[{{Garro} {et~al.}(2020){Garro}, {Minniti}, {G{\'o}mez},
  {Alonso-Garc{\'\i}a}, {Barb{\'a}}, {Barbuy}, {Clari{\'a}}, {Chen{\'e}},
  {Dias}, {Hempel}, {Ivanov}, {Lucas}, {Majaess}, {Mauro}, {Moni Bidin},
  {Palma}, {Pullen}, {Saito}, {Smith}, {Surot}, {Ram{\'\i}rez Alegr{\'\i}a},
  {Rejkuba}, {Ripepi}, \& {Fern{\'a}ndez Trincado}}]{2020A&A...642L..19G}
{Garro}, E.~R., {Minniti}, D., {G{\'o}mez}, M., {et~al.} 2020, \aap, 642, L19

\bibitem[{{Garro} {et~al.}(2022{\natexlab{b}}){Garro}, {Minniti}, {G{\'o}mez},
  {Fern{\'a}ndez-Trincado}, {Alonso-Garc{\'\i}a}, {Hempel}, \& {Zelada
  Bacigalupo}}]{2022A&A...662A..95G}
{Garro}, E.~R., {Minniti}, D., {G{\'o}mez}, M., {et~al.} 2022{\natexlab{b}},
  \aap, 662, A95

\bibitem[{{Gonz{\'a}lez-Fern{\'a}ndez}
  {et~al.}(2018){Gonz{\'a}lez-Fern{\'a}ndez}, {Hodgkin}, {Irwin},
  {Gonz{\'a}lez-Solares}, {Koposov}, {Lewis}, {Emerson}, {Hewett},
  {Yolda{\c{s}}}, \& {Riello}}]{2018MNRAS.474.5459G}
{Gonz{\'a}lez-Fern{\'a}ndez}, C., {Hodgkin}, S.~T., {Irwin}, M.~J., {et~al.}
  2018, \mnras, 474, 5459

\bibitem[{{Griggio} {et~al.}(2022){Griggio}, {Bedin}, {Raddi}, {Reindl},
  {Tomasella}, {Scalco}, {Salaris}, {Cassisi}, {Ochner}, {Ciroi}, {Rosati},
  {Nardiello}, {Anderson}, {Libralato}, {Bellini}, {Vallenari}, {Spina}, \&
  {Pedani}}]{2022MNRAS.515.1841G}
{Griggio}, M., {Bedin}, L.~R., {Raddi}, R., {et~al.} 2022, \mnras, 515, 1841

\bibitem[{{Griggio} {et~al.}(2023){Griggio}, {Salaris}, {Nardiello}, {Bedin},
  {Cassisi}, \& {Anderson}}]{2023MNRAS.524..108G}
{Griggio}, M., {Salaris}, M., {Nardiello}, D., {et~al.} 2023, \mnras, 524, 108

\bibitem[{{H{\"a}berle} {et~al.}(2021){H{\"a}berle}, {Libralato}, {Bellini},
  {Watkins}, {Pott}, {Neumayer}, {van der Marel}, {Piotto}, \&
  {Nardiello}}]{2021MNRAS.503.1490H}
{H{\"a}berle}, M., {Libralato}, M., {Bellini}, A., {et~al.} 2021, \mnras, 503,
  1490

\bibitem[{{Hajdu} {et~al.}(2020){Hajdu}, {D{\'e}k{\'a}ny}, {Catelan}, \&
  {Grebel}}]{2020ExA....49..217H}
{Hajdu}, G., {D{\'e}k{\'a}ny}, I., {Catelan}, M., \& {Grebel}, E.~K. 2020,
  Experimental Astronomy, 49, 217

\bibitem[{{Irwin} {et~al.}(2004){Irwin}, {Lewis}, {Hodgkin}, {Bunclark},
  {Evans}, {McMahon}, {Emerson}, {Stewart}, \& {Beard}}]{2004SPIE.5493..411I}
{Irwin}, M.~J., {Lewis}, J., {Hodgkin}, S., {et~al.} 2004, in Society of
  Photo-Optical Instrumentation Engineers (SPIE) Conference Series, Vol. 5493,
  Optimizing Scientific Return for Astronomy through Information Technologies,
  ed. P.~J. {Quinn} \& A.~{Bridger}, 411--422

\bibitem[{{Kader} {et~al.}(2022){Kader}, {Pilachowski}, {Johnson}, {Rich},
  {Young}, {Simion}, {Clarkson}, {Michael}, {Kunder}, {Vivas}, {Koch-Hansen},
  \& {Marchetti}}]{2022ApJ...940...76K}
{Kader}, J.~A., {Pilachowski}, C.~A., {Johnson}, C.~I., {et~al.} 2022, \apj,
  940, 76

\bibitem[{{Lewis} {et~al.}(2010){Lewis}, {Irwin}, \&
  {Bunclark}}]{2010ASPC..434...91L}
{Lewis}, J.~R., {Irwin}, M., \& {Bunclark}, P. 2010, in Astronomical Society of
  the Pacific Conference Series, Vol. 434, Astronomical Data Analysis Software
  and Systems XIX, ed. Y.~{Mizumoto}, K.~I. {Morita}, \& M.~{Ohishi}, 91

\bibitem[{{Libralato} {et~al.}(2015){Libralato}, {Bellini}, {Bedin},
  {Anderson}, {Piotto}, {Nascimbeni}, {Platais}, {Minniti}, \&
  {Zoccali}}]{libra2015}
{Libralato}, M., {Bellini}, A., {Bedin}, L.~R., {et~al.} 2015, \mnras, 450,
  1664

\bibitem[{{Libralato} {et~al.}(2014){Libralato}, {Bellini}, {Bedin}, {Piotto},
  {Platais}, {Kissler-Patig}, \& {Milone}}]{2014A&A...563A..80L}
{Libralato}, M., {Bellini}, A., {Bedin}, L.~R., {et~al.} 2014, \aap, 563, A80

\bibitem[{{Libralato} {et~al.}(2021){Libralato}, {Lennon}, {Bellini}, {van der
  Marel}, {Clark}, {Najarro}, {Patrick}, {Anderson}, {Bedin}, {Crowther}, {de
  Mink}, {Evans}, {Platais}, {Sabbi}, \& {Sohn}}]{2021MNRAS.500.3213L}
{Libralato}, M., {Lennon}, D.~J., {Bellini}, A., {et~al.} 2021, \mnras, 500,
  3213

\bibitem[{{Minniti}(2018)}]{2018ASSP...51...63M}
{Minniti}, D. 2018, in Astrophysics and Space Science Proceedings, Vol.~51, The
  Vatican Observatory, Castel Gandolfo: 80th Anniversary Celebration, ed.
  G.~{Gionti} \& J.-B. {Kikwaya Eluo}, 63

\bibitem[{{Minniti} {et~al.}(2021{\natexlab{a}}){Minniti},
  {Fern{\'a}ndez-Trincado}, {G{\'o}mez}, {Smith}, {Lucas}, \& {Contreras
  Ramos}}]{2021A&A...650L..11M}
{Minniti}, D., {Fern{\'a}ndez-Trincado}, J.~G., {G{\'o}mez}, M., {et~al.}
  2021{\natexlab{a}}, \aap, 650, L11

\bibitem[{{Minniti} {et~al.}(2021{\natexlab{b}}){Minniti},
  {Fern{\'a}ndez-Trincado}, {Smith}, {Lucas}, {G{\'o}mez}, \&
  {Pullen}}]{2021A&A...648A..86M}
{Minniti}, D., {Fern{\'a}ndez-Trincado}, J.~G., {Smith}, L.~C., {et~al.}
  2021{\natexlab{b}}, \aap, 648, A86

\bibitem[{{Minniti} {et~al.}(2017){Minniti}, {Geisler}, {Alonso-Garc{\'\i}a},
  {Palma}, {Beam{\'\i}n}, {Borissova}, {Catelan}, {Clari{\'a}}, {Cohen},
  {Contreras Ramos}, {Dias}, {Fern{\'a}ndez-Trincado}, {G{\'o}mez}, {Hempel},
  {Ivanov}, {Kurtev}, {Lucas}, {Moni-Bidin}, {Pullen}, {Ram{\'\i}rez
  Alegr{\'\i}a}, {Saito}, \& {Valenti}}]{2017ApJ...849L..24M}
{Minniti}, D., {Geisler}, D., {Alonso-Garc{\'\i}a}, J., {et~al.} 2017, \apjl,
  849, L24

\bibitem[{{Minniti} {et~al.}(2010){Minniti}, {Lucas}, {Emerson}, {Saito},
  {Hempel}, {Pietrukowicz}, {Ahumada}, {Alonso}, {Alonso-Garcia}, {Arias},
  {Bandyopadhyay}, {Barb{\'a}}, {Barbuy}, {Bedin}, {Bica}, {Borissova},
  {Bronfman}, {Carraro}, {Catelan}, {Clari{\'a}}, {Cross}, {de Grijs},
  {D{\'e}k{\'a}ny}, {Drew}, {Fari{\~n}a}, {Feinstein}, {Fern{\'a}ndez
  Laj{\'u}s}, {Gamen}, {Geisler}, {Gieren}, {Goldman}, {Gonzalez}, {Gunthardt},
  {Gurovich}, {Hambly}, {Irwin}, {Ivanov}, {Jord{\'a}n}, {Kerins}, {Kinemuchi},
  {Kurtev}, {L{\'o}pez-Corredoira}, {Maccarone}, {Masetti}, {Merlo},
  {Messineo}, {Mirabel}, {Monaco}, {Morelli}, {Padilla}, {Palma}, {Parisi},
  {Pignata}, {Rejkuba}, {Roman-Lopes}, {Sale}, {Schreiber}, {Schr{\"o}der},
  {Smith}, {}, {Soto}, {Tamura}, {Tappert}, {Thompson}, {Toledo}, {Zoccali}, \&
  {Pietrzynski}}]{vvv2010}
{Minniti}, D., {Lucas}, P.~W., {Emerson}, J.~P., {et~al.} 2010, \na, 15, 433

\bibitem[{{Minniti} {et~al.}(2023){Minniti}, {Matsunaga}, {Fernandez-Trincado},
  {Otsubo}, {Sarugaku}, {Takeuchi}, {Katoh}, {Hamano}, {Ikeda}, {Kawakita},
  {Lucas}, {Smith}, {Petralia}, {Garro}, {Saito}, {Alonso-Garcia}, {Gomez}, \&
  {Navarro}}]{2023arXiv231216028M}
{Minniti}, D., {Matsunaga}, N., {Fernandez-Trincado}, J.~G., {et~al.} 2023,
  arXiv e-prints, arXiv:2312.16028

\bibitem[{{Mr{\'o}z} {et~al.}(2019){Mr{\'o}z}, {Udalski}, {Skowron},
  {Szyma{\'n}ski}, {Soszy{\'n}ski}, {Wyrzykowski}, {Pietrukowicz},
  {Koz{\l}owski}, {Poleski}, {Ulaczyk}, {Rybicki}, \&
  {Iwanek}}]{2019ApJS..244...29M}
{Mr{\'o}z}, P., {Udalski}, A., {Skowron}, J., {et~al.} 2019, \apjs, 244, 29

\bibitem[{{Platais} {et~al.}(2002){Platais}, {Kozhurina-Platais}, {Girard},
  {van Altena}, {Klemola}, {Stauffer}, {Armandroff}, {Mighell}, {Dell'Antonio},
  {Falco}, \& {Sarajedini}}]{2002AJ....124..601P}
{Platais}, I., {Kozhurina-Platais}, V., {Girard}, T.~M., {et~al.} 2002, \aj,
  124, 601

\bibitem[{{Platais} {et~al.}(2006){Platais}, {Wyse}, \&
  {Zacharias}}]{2006PASP..118..107P}
{Platais}, I., {Wyse}, R. F.~G., \& {Zacharias}, N. 2006, \pasp, 118, 107

\bibitem[{{Scalco} {et~al.}(2021){Scalco}, {Bellini}, {Bedin}, {Anderson},
  {Rosati}, {Libralato}, {Salaris}, {Vesperini}, {Nardiello}, {Apai},
  {Burgasser}, \& {Gerasimov}}]{2021MNRAS.505.3549S}
{Scalco}, M., {Bellini}, A., {Bedin}, L.~R., {et~al.} 2021, \mnras, 505, 3549

\bibitem[{{Skrutskie} {et~al.}(2006){Skrutskie}, {Cutri}, {Stiening},
  {Weinberg}, {Schneider}, {Carpenter}, {Beichman}, {Capps}, {Chester},
  {Elias}, {Huchra}, {Liebert}, {Lonsdale}, {Monet}, {Price}, {Seitzer},
  {Jarrett}, {Kirkpatrick}, {Gizis}, {Howard}, {Evans}, {Fowler}, {Fullmer},
  {Hurt}, {Light}, {Kopan}, {Marsh}, {McCallon}, {Tam}, {Van Dyk}, \&
  {Wheelock}}]{2006AJ....131.1163S}
{Skrutskie}, M.~F., {Cutri}, R.~M., {Stiening}, R., {et~al.} 2006, \aj, 131,
  1163

\bibitem[{{Smith} {et~al.}(2018){Smith}, {Lucas}, {Kurtev}, {Smart}, {Minniti},
  {Borissova}, {Jones}, {Zhang}, {Marocco}, {Contreras Pe{\~n}a}, {Gromadzki},
  {Kuhn}, {Drew}, {Pinfield}, \& {Bedin}}]{virac}
{Smith}, L.~C., {Lucas}, P.~W., {Kurtev}, R., {et~al.} 2018, \mnras, 474, 1826

\bibitem[{{Zoccali}(2019)}]{2019BAAA...61..137Z}
{Zoccali}, M. 2019, Boletin de la Asociacion Argentina de Astronomia La Plata
  Argentina, 61, 137

\end{thebibliography}

\end{document}